\documentclass{aa}  
\usepackage{graphicx}
\usepackage{txfonts}

\newcommand{\ros}{{\it ROSAT}}

\newcommand{\xmm}{{\it XMM-Newton}}
\newcommand{\eros}{{eROSITA}}
\newcommand{\nicer}{{\it NICER}}
\newcommand{\swift}{{\it Swift}}
\newcommand{\cxo}{{\it Chandra}}

\newcommand{\lbtl}{{Large Binocular Telescope}}
\newcommand{\lbt}{{LBT}}
\newcommand{\saltl}{{Southern African Large Telescope}}
\newcommand{\salt}{{SALT}}
\newcommand{\nh}{N_{\rm H}}
\newcommand{\gaia}{{\it GAIA}}

\def \magoe{\object{RX~J1856.5-3754}}

\def \magzf{\object{RX~J0420.0-5022}}
\def \carINS{\object{2XMM~J104608.7-594306}}

\def \zsfs{\object{J0657}}
\def \otos{\object{J1317}}
\def \tmzsfs{\object{PSR~B0656+14}}
\def \tmgem{\object{Geminga}}
\def \tmozff{\object{PSR~B1055-52}}

\def \ztto{\object{4XMM J022141.5-735632}}
\def \fluxcgs{erg~s$^{-1}$~cm$^{-2}$}
\def \jotos{\object{eRASSU~J131716.9-402647}} 
\def \jzsfs{\object{eRASSU~J065715.3+260428}}
\def \jotoss{\object{J131716.9-402647}} 
\def \jzsfss{\object{J065715.3+260428}}

\begin{document} 

\title{Discovery of two promising isolated neutron star candidates in the SRG/eROSITA All-Sky Survey} 
\author{J.~Kurpas\inst{1,2}
   \and A.~D.~Schwope\inst{1}
   \and A.~M.~Pires\inst{1}
   \and F.~Haberl\inst{3}
   \and D.~A.~H.~Buckley\inst{4,5,6}
}
\offprints{J. Kurpas}
\institute{Leibniz-Institut f\"ur Astrophysik Potsdam (AIP), An der Sternwarte 16, 14482 Potsdam, Germany
   \email{jkurpas@aip.de} 
   \and
   Potsdam University, Institute for Physics and Astronomy, Karl-Liebknecht-Stra\ss e 24/25, 14476 Potsdam, Germany
   \and
   Max-Planck-Institut f\"ur extraterrestrische Physik, Giessenbachstra\ss e, 85748 Garching, Germany
\and
South African Astronomical Observatory, PO Box 9, Observatory 7935, Cape Town, South Africa
\and
Department of Physics, University of the Free State, PO Box 339, Bloemfontein 9300, South Africa
\and
Department of Astronomy, University of Cape Town, Private Bag X3, Rondebosch 7701, South Africa
}
\date{Received ...; accepted ...}
\titlerunning{Two isolated neutron star candidates from the eRASS}
\authorrunning{J.~Kurpas et al.}
\abstract
{We report the discovery of the isolated neutron star (INS) candidates \jzsfs\ and \jotos\ from the {\it Spectrum Roentgen Gamma} ({\it SRG}) eROSITA All-Sky Survey. Selected for their soft X-ray emission and absence of catalogued counterparts, both objects were recently targeted with the \lbtl\ and the \saltl. The absence of counterparts down to deep optical limits (25 mag, 5$\sigma$) and, as a result, large X-ray-to-optical flux ratios in both cases strongly suggest an INS nature. The X-ray spectra of both sources are well described by a simple absorbed blackbody, whereas other thermal and non-thermal models (e.g. a hot-plasma emission spectrum or power law) are disfavoured by the spectral analysis. Within the current observational limits, and as expected for cooling INSs, no significant variation ($>2\sigma$) has been identified over the first two-year time span of the survey. Upcoming dedicated follow-up observations will help us to confirm the candidates' nature.}
\maketitle
\section{Introduction\label{sec_intro}}

The majority of the known Galactic isolated neutron star (INS) population consists of radio and $\gamma$-ray pulsars and recycled millisecond pulsars. Powered by the loss of rotational energy \citep{2009ASSL..357...91B}, these objects are preferentially detected in large-scale radio surveys. More elusive are 'magnetars' \citep[see, e.g. ][for a review]{2017ARA&A..55..261K}, which are driven by their super strong magnetic energy; the young and atypical central compact objects (CCOs) in supernova remnants \citep{2017JPhCS.932a2006D}; and the radio-quiet, so-called X-ray dim isolated neutron stars \cite[XDINSs,][]{2009ASSL..357..141T}. In particular, XDINSs are well known for their purely thermal spectra, with no significant traces of magnetospheric or accretion activity. Only recently were weak non-thermal emission components detected in the cumulative X-ray spectra of the XDINS \magoe\ and \magzf\ \citep{2020ApJ...904...42D, 2022MNRAS.516.4932D}. XDINSs possess spin periods ranging from 3--17~s, which, under the assumption of magneto-dipole radiation being solely responsible for the observed braking of the neutron star rotation \citep{1969ApJ...157.1395O}, imply spin-down ages on the million year timescale. Nevertheless, XDINSs are powered by thermal emission and not magnetic braking, and thus kinematic age estimates of the order of $0.5-1$~Myr \citep{2010MNRAS.402.2369T, 2011MNRAS.417..617T, 2012PASA...29...98T} are deemed more reliable, whereas statistical age estimates suffer from the small known population \citep{2013MNRAS.435.3243G}.  In comparison to similarly aged rotation-powered pulsars, XDINSs possess larger spin periods, stronger magnetic field strengths ($10^{13}-10^{14}$~G), and higher thermal luminosities.

Originally discovered in \ros\ observations \citep[see][for details]{2007Ap&SS.308..181H}, the seven known XDINSs are rather nearby and could be as common as rotation-powered pulsars. \cite{2008MNRAS.391.2009K} estimated that the total INS birthrate exceeds the Galactic core-collapse supernova rate, which might imply evolutionary connections between the different classes of INSs. This is supported by state-of-the art magnetothermal evolutionary models that suggest XDINSs are evolved magnetars \citep{2013MNRAS.434..123V}. The small number of confirmed XDINSs makes it difficult to investigate population properties on the Galactic scale and to confirm evolutionary links with other INS families. Thus, attempts have been made since their initial discovery to find new members. Most notable are the identification of Calvera \citep{2008ApJ...672.1137R}, a fast-spinning INS whose exact nature is still under debate \citep{2019ApJ...877...69B}; \carINS,\ a seemingly cooling neutron star in the Carina Nebula \citep{2009A&A...504..185P, 2015A&A...583A.117P}; and the four INS candidates recently selected from the \xmm\ footprint \citep{2022MNRAS.509.1217R}. Among these, \ztto\ appears particularly promising on the basis of its purely thermal X-ray emission and stable properties \citep{2022A&A...666A.148P}. 

\begin{table*}[t]
\caption{Properties of the two targets from eRASS:4.
\label{tab_erass_prop}}
\centering
\begin{tabular}{lrrccrrrrr}
\hline\hline
Target & RA & Dec & Error\,\tablefootmark{(a)} & $N_{\rm H}^{\rm Gal}$\,\tablefootmark{(b)} & $b$ & Flux\,\tablefootmark{(c)} & HR$_1$\,\tablefootmark{(d)} & HR$_2$\,\tablefootmark{(d)} & HR$_3$\,\tablefootmark{(d)}\\
{eRASSU}  & [\degr] & [\degr] & [\arcsec] & [$10^{20}$~cm$^{-2}$] & [\degr] & & & & \\
\hline
\jzsfss & $104.3140$ & $26.0746$ & $1.3$ & $8.3$ & 12.75 & $2.21(26)$ & $0.33(11)$ & $-0.79(9)$ & $-1.00(26)$\\ 
\jotoss & $199.3204$ & $-40.4462$ & $0.9$ & $6.7$ & 22.16 & $4.52(28)$ & $0.42(6)$ & $-0.83(4)$ & $-0.92(18)$\\ 
\hline
\end{tabular}
\tablefoot{
\tablefoottext{a}{The listed positional accuracy was computed adding the combined statistical 1$\sigma$ positional error ($r_\mathrm{c}$), estimated from maximum likelihood point spread function fitting \citep{2022A&A...661A...1B}, with a systematic error of 0.6\arcsec and a stretching factor of 1.1 via the equation $r_\mathrm{pos} = \sqrt{\left(r_\mathrm{c}\times 1.1/\sqrt{2}\right)^2+\left(0.6\arcsec\right)^2}$. The later two values were estimated by cross-matching the eRASS:4 catalogue with the 'astrometric selection' sample derived from the \gaia\ DR3 qso\_candidates table \citep{2022arXiv220605681G} and fitting the offset distribution with a Rayleigh function.}
\tablefoottext{b}{Galactic line-of-sight hydrogen column density \citep{2016A&A...594A.116H}.}
\tablefoottext{c}{The stacked eRASS:4 flux is given in units of $10^{-13}$~\fluxcgs\ in the $0.2-2$~keV energy band}
\tablefoottext{d}{Hardness ratios (HRs) compare the count difference to the total number of counts in adjacent energy bands. The bands range from $0.2-0.5$~keV, $0.5-1$~keV, $1-2$~keV, and $2-5$~keV.}}
\end{table*}

The discovery of additional XDINS candidates beyond the solar vicinity has been met with difficulties in the past. While the \ros\ survey \citep[][]{1999A&A...349..389V} covers the full sky, the large positional errors at faint X-ray fluxes make the selection of new candidates impractical due to source confusion. X-ray catalogues based on instruments that offer a better localisation and often reach fainter fluxes (e.g. \xmm\ or \cxo) are limited by the small sky coverage. The \eros\ All-Sky Survey \citep[eRASS,][]{2021A&A...647A...1P} is conducting the deepest imaging of the entire X-ray sky since the days of \ros. At soft X-ray energies, eRASS ought to surpass its predecessor by a factor of 20 in sensitivity \citep{2012arXiv1209.3114M}, with a much improved spectral resolution and positional accuracy. Flux-limited searches over the German part of the \eros\ sky\footnote{The \eros\ data rights are equally split between a German and a Russian consortium; see \citet{2021A&A...647A...1P} for details.} thence offer the unique opportunity to unveil their numbers and birthrate beyond the immediate solar vicinity \citep{2017AN....338..213P}. To this end, we searched the eRASS catalogues\footnote{Created by and made available to the members of the German \eros\ consortium} for X-ray sources with properties consistent with those of INSs. We followed the brightest among our candidates up with the \lbtl\ \citep[LBT,][]{2012SPIE.8444E..1AH} and \saltl\ \citep[SALT,][]{2006SPIE.6267E..0ZB} to rule out other soft X-ray emitters that are known to contaminate our sample. We present in this work the first two promising INS candidates resulting from our ongoing identification campaign, \jzsfs\ and \jotos\ (hereafter dubbed \zsfs\ and \otos). 

The paper is structured as follows. In Sect.~\ref{sec_obs} we present the applied search strategy and the available observations at optical and X-ray energies. We present the properties inferred from the X-ray spectra and the obtained optical magnitude limits in Sect.~\ref{sec_analysis} and we discuss the results in Sect.~\ref{sec_disc}.

\section{Observations and data reduction\label{sec_obs}}

\begin{table}
\caption{X-ray and optical observations of \zsfs\ and \otos.
\label{tab_obs}}
\centering

\begin{tabular}{lllrrrrrr}
\hline\hline
Target & Telescope & Date of & Cts\,\tablefootmark{(a)} & T$_\mathrm{exp}$\,\tablefootmark{(b)} \\
 & & Observation & & [s]\\
\hline
\zsfs & \eros & 2020-04-11    & 21 & $126$\\
\zsfs & \eros & 2020-10-14/15 & 19 & $130$\\
\zsfs & \eros & 2021-04-10/11 & 30 & $157$\\
\zsfs & \eros & 2021-10-12/13 & 26 & $185$\\
\hline
\otos & \eros & 2020-01-19/20/21 & 107 & $353$\\
\otos & \eros & 2020-07-21/22/23 & 104 & $345$\\
\otos & \eros & 2021-01-15/16 & 75 & $222$\\
\otos & \eros & 2021-07-23/24 & 70 & $230$\\
\otos & \eros & 2022-01-21/22 & 62 & $247$\\
\hline
\zsfs & \lbt & 2021-12-05 & & $400$\\
\zsfs & \lbt & 2022-04-01 & & $400$\\
\hline
\otos & \salt & 2022-05-26 & & $2400$\\
\hline
\end{tabular}
\tablefoot{During the observations eROSITA was operated in SURVEY mode.
\tablefoottext{a}{Number of detected counts in the 0.2--5.0~keV band}
\tablefoottext{b}{The given X-ray exposure times do not account for vignetting, but state the CCD live time only.}}
\end{table}

\subsection{Target selection}
Since the start of the survey observations in December 2019, \eros\ has completed four out of the eight planned full sky scans and started the fifth. We searched the single scan and stacked (which combines all four completed surveys, dubbed eRASS:4) catalogues for possible candidates above flux values of $10^{-13}$\fluxcgs\ (0.2--2.0~keV). The eRASS will contain fainter INSs, but those objects are difficult to separate from other soft X-ray emitting sources based on the currently available positional accuracy alone. At the same time, the necessary optical follow-up observations will be expensive (see \cite{2017AN....338..213P,2021ARep...65..615K} for the prospect of INS searches in eRASS), so the confirmation of fainter candidates is best left to the next generation of large telescopes (e.g. ESO-ELT) and/or sky surveys \citep[e.g. the LSST, ][]{2019ApJ...873..111I}.

\begin{figure}[t]
\begin{center}
\includegraphics[width=\linewidth]{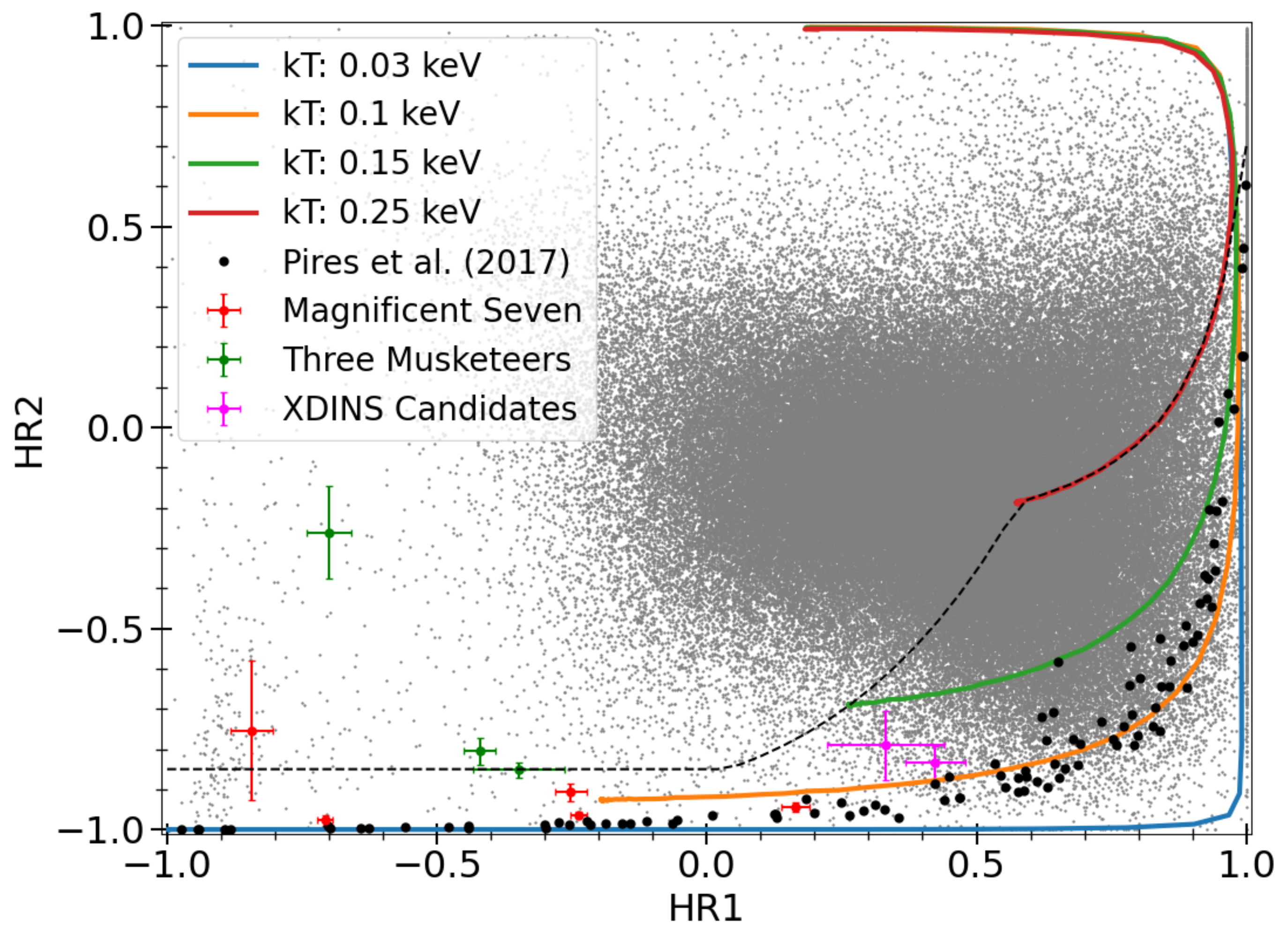}\vskip1pt
\includegraphics[width=\linewidth]{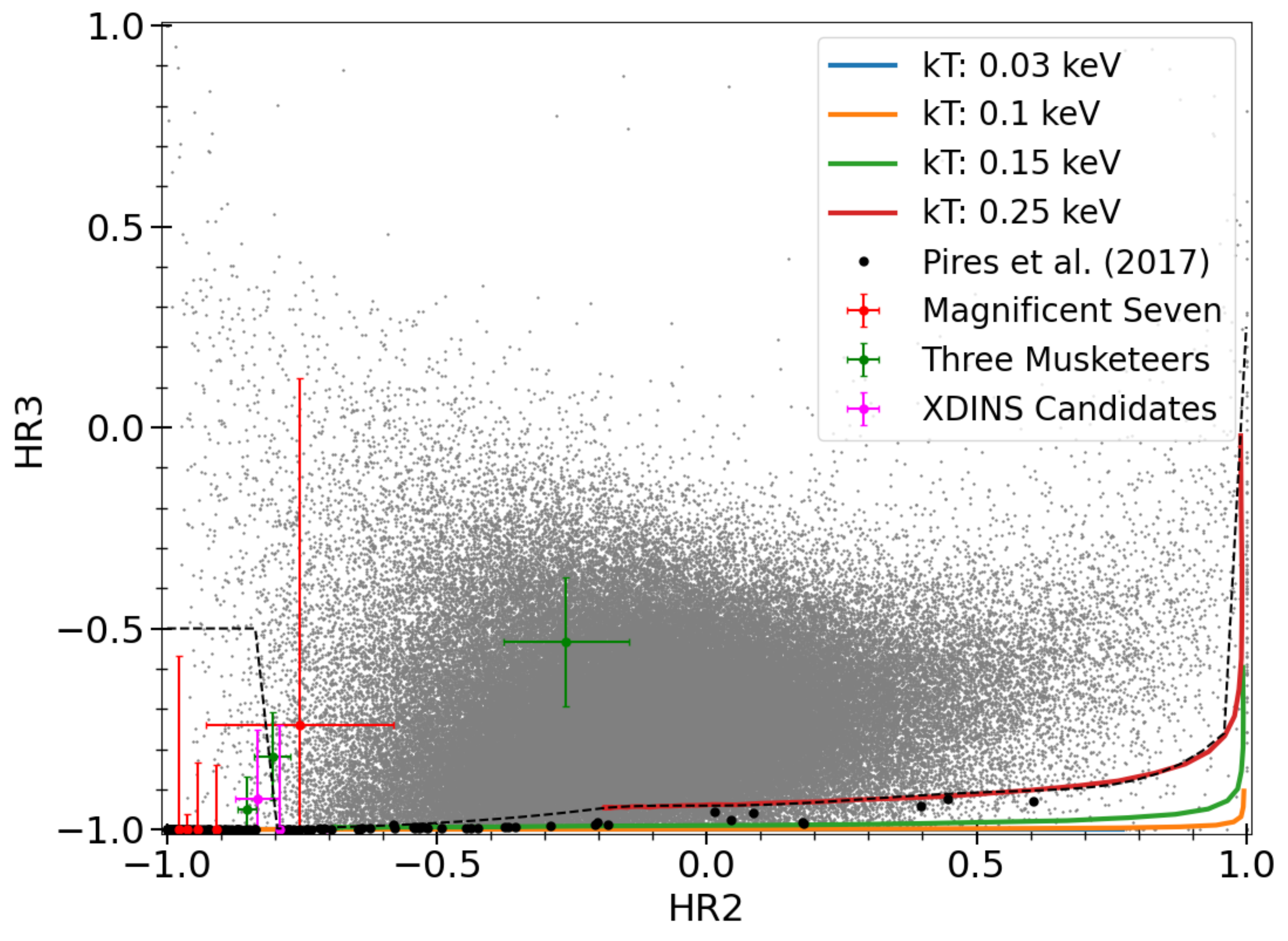}\vskip1pt
\end{center}
\caption{Hardness ratio diagrams indicating the applied selection (dashed black
 line), the distribution of simulated absorbed blackbody spectra (solid lines), and the position of the synthetic thermally emitting INS population of \cite{2017AN....338..213P} (black circles). The known population of XDINSs (dubbed the `magnificent seven'), on the German part of the eROSITA sky, in red, and the three middle-aged thermally emitting rotation-powered pulsars, \tmzsfs, \tmgem, and \tmozff\ (dubbed the `three musketeers') in green are indicated along with the location of the two INS candidates discussed in this work (magenta). The hardness ratios were computed between the 0.2--0.5~keV and 0.5--1.0~keV (HR1), 0.5--1.0~keV and 1.0--2.0~keV (HR2), and 1.0--2.0~keV and 2.0--5.0~keV (HR3) bands.}
\label{fig_hr_cut}
\end{figure}

The known XDINSs are characterised by their soft X-ray spectra, which are well described by absorbed blackbody emission, with effective temperatures between 40--100~eV \citep{2007Ap&SS.308..181H}. Using counts in neighbouring energy bands to compute hardness ratios, the soft nature can be translated into hardness ratio constraints. The applied cuts (using hardness ratios computed between the 0.2--0.5~keV and 0.5--1~keV, the 0.5--1~keV and 1--2~keV, and the 1--2~keV and 2--5~keV bands) were defined based on XSPEC \citep[][version: 12.12.0]{1996ASPC..101...17A} simulations of absorbed blackbody spectra with $\nh$ values ranging from $0-10^{23}$~cm$^{-2}$ and effective temperatures between 30--250~eV. The resulting cuts are shown in Fig.~\ref{fig_hr_cut}.

XDINSs possess X-ray-to-optical flux ratio values of $\log(f_\mathrm{X}/f_{\mathrm{opt,V}})\geq 4.0$ \citep{2003ApJ...588L..33K}. Thus, at the X-ray flux level chosen, optical counterparts are expected to be fainter than magnitude 29, which is too faint to be detected in publicly available wide-area surveys and databases (e.g. the 5$\sigma$ r-band detection limits of the deepest used optical surveys in this work, namely Pan-STARRS DR1 \citep{2016arXiv161205560C} and Legacy Survey DR9/10 \citep{2019AJ....157..168D}, are located at 23.2~mag and 24--25.3~mag, respectively). This allows one to select candidates by requiring for the absence of optical/infrared (IR) counterparts. Specifically, we conducted a probabilistic cross-matching of the eRASS position against the \gaia\ DR3 \citep{2022arXiv220800211G}, Pan-STARRS DR1, and Legacy Survey DR9/10 catalogues using the ARCHES X-Match tool \citep{2017A&A...597A..89P}. We kept candidates with non-matching probabilities in excess of 50\%. Additionally, we excluded sources with counterparts in the SIMBAD database \citep{2000A&AS..143....9W}, the VISTA Hemisphere Survey \citep{2013Msngr.154...35M}, and the CatWISE2020 \citep{2021ApJS..253....8M} catalogues.

To exclude other potential identifications -- for example X-ray emitting stars, active galactic nuclei (AGN), or cataclysmic variables --, additional observations at optical energies are necessary. Thus, we initiated a follow-up campaign with \lbt\ and \salt, to confirm the absence of counterparts at fainter magnitudes. While this observational campaign is still ongoing, optical imaging has already been conducted for \zsfs\ and \otos\ (Table~\ref{sec_obs}). A paper reporting on the full search results in the eRASS catalogues will be presented in another work (Kurpas et al., in preparation).


\begin{figure*}[t]
\begin{minipage}{0.49\textwidth}
\begin{center}
\includegraphics[width=\linewidth]{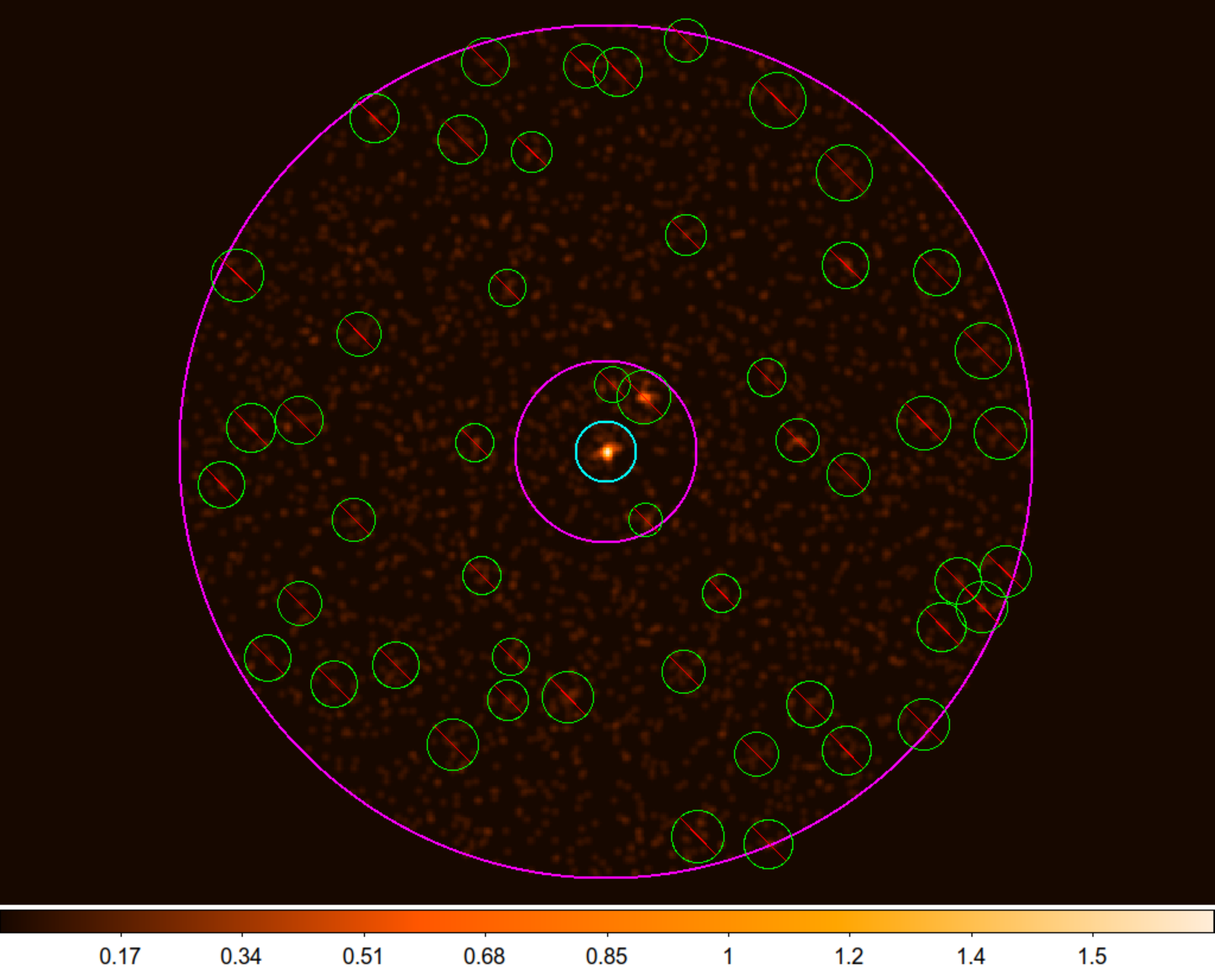}\vskip1pt
\end{center}
\end{minipage}\hspace*{2.5mm}
\begin{minipage}{0.49\textwidth}
\includegraphics[width=\linewidth]{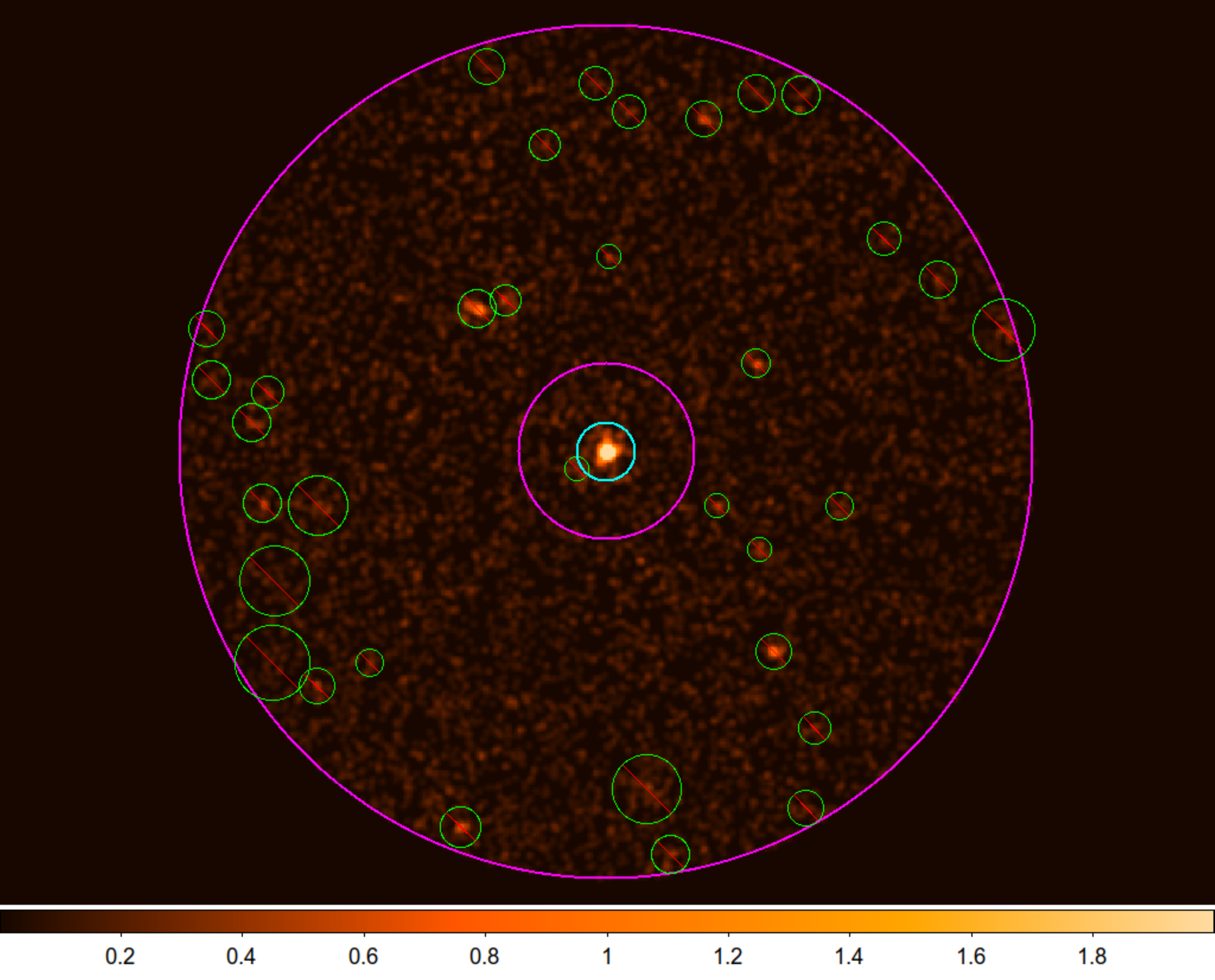}
\end{minipage}
\caption{Stacked eRASS source (cyan circle, 1\arcmin\ radius) and background (magenta annulus, 3\arcmin\ inner and 15\arcmin\ outer radius) regions for \zsfs\ (left) and \otos\ (right). Field sources discovered from a source detection run, using all telescopes (TM0) and photons in the 0.2--10~keV band, were excluded from the source and background region (green circles). The images include photons from the same telescope and energy band configuration that was used for the source detection.}
\label{fig_reg}
\end{figure*}



\begin{table*}
\caption{X-ray spectral fitting results for both candidates.
\label{tab_specfit}}
\centering

\begin{tabular}{ccccccccc}
\hline\hline
\multicolumn{5}{l}{\texttt{BB}}\\
\hline
Source & $\nh$ & $kT$ & Radius\tablefootmark{(a)} & log(Z) & log(Z$_\mathrm{BB}$/Z) & AIC & Absorbed flux\tablefootmark{(b)}\\
 & [$10^{20}$~cm$^{-2}$] & [eV] & [km] & & & & [$10^{-13}$~\fluxcgs]\\ 
\hline
\zsfs & $5^{+5}_{-4}$ & $108^{+15}_{-14}$ & $1.9^{+1.4}_{-0.7}$ &  -37.4 & 0 & 65.7 & $2.29^{+0.29}_{-0.26}$ \\ 
\otos & $10.2^{+3}_{-2.8}$ & $110^{+7}_{-7}$ & $3.0^{+0.9}_{-0.7}$ &  -71.2 & 0 & 128.5 & $4.40^{+0.25}_{-0.24}$ \\ 
\hline
\multicolumn{5}{l}{\texttt{apec}\tablefootmark{(c)}}\\
\hline
Source & $\nh$ & $kT$ & $z$  & log(Z) & log(Z$_\mathrm{BB}$/Z) & AIC & Absorbed flux\tablefootmark{(b)}\\
 & [$10^{20}$~cm$^{-2}$] & [eV] & & & & & [$10^{-13}$~\fluxcgs]\\ 
\hline
\zsfs & $1.9^{+2.6}_{-1.8}$ & $203^{+17}_{-19}$ & $0$ & -39.5 & 2.0  & 68.0 & $2.35^{+0.3}_{-0.27}$ \\ 
\otos & $3.0^{+1.9}_{-1.8}$ & $281^{+15}_{-14}$ & $0.170^{+0.016}_{-0.022}$ &  -72.1 & 0.9 & 118.9 & $4.56^{+0.25}_{-0.25}$ \\ 
\hline
\multicolumn{5}{l}{\texttt{PL}}\\
\hline
Source & $\nh$ & $\Gamma$ &  & log(Z) & log(Z$_\mathrm{BB}$/Z) & AIC & Absorbed flux\tablefootmark{(b)}\\
 & [$10^{20}$~cm$^{-2}$] & & & & & & [$10^{-13}$~\fluxcgs]\\ 
\hline
\zsfs & $17^{+6}_{-5}$ & $6.5^{+0.9}_{-0.7}$ & &  -40.3 & 2.9 & 68.9 & $2.31^{+0.29}_{-0.26}$ \\ 
\otos & $31^{+5}_{-5}$ & $7.3^{+0.5}_{-0.5}$ & &  -83.7 & 12.5 & 151.2 & $4.39^{+0.22}_{-0.25}$ \\ 
\hline
\multicolumn{5}{l}{\texttt{NSA}\tablefootmark{(d)}}\\
\hline
Source & $\nh$ & $kT$ & Magn. Field & Dist. & log(Z$_\mathrm{BB}$/Z) & AIC & Absorbed flux\tablefootmark{(b)}\\
 & [$10^{20}$~cm$^{-2}$] & [eV] & [G] & [pc] & & & [$10^{-13}$~\fluxcgs]\\ 
\hline
\zsfs & $10^{+4}_{-4}$ & $31^{+8}_{-6}$ & $0$       & $230^{+280}_{-130}$ & 1.4 &  67.1 & $2.28^{+0.29}_{-0.27}$ \\ 
\zsfs & $11^{+5}_{-3}$ & $46^{+7}_{-8}$ & $10^{12}$ & $480^{+400}_{-260}$ & 1.2  &  66.6 & $2.26^{+0.27}_{-0.26}$ \\ 
\zsfs & $11^{+5}_{-3}$ & $49^{+6}_{-8}$ & $10^{13}$ & $550^{+300}_{-270}$ & 1.5  &  66.4 & $2.25^{+0.28}_{-0.26}$ \\ 
\otos & $16^{+4}_{-4}$ & $31^{+4}_{-4}$ & $0$       & $130^{+ 80}_{- 50}$ & 4.1  & 135.1 & $4.38^{+0.26}_{-0.24}$ \\
\otos & $16^{+4}_{-4}$ & $49^{+6}_{-6}$ & $10^{12}$ & $340^{+190}_{-130}$ & 2.9  & 133.4 & $4.39^{+0.24}_{-0.28}$ \\
\otos & $17^{+4}_{-4}$ & $52^{+6}_{-6}$ & $10^{13}$ & $400^{+200}_{-150}$ & 3.1  & 133.3 & $4.38^{+0.26}_{-0.24}$ \\
\hline
\end{tabular}
\tablefoot{We present the median value estimated from the sample parameter distribution and the errors give the 1$\sigma$ confidence region. We adopted 400 live points to sample the parameter space.
\tablefoottext{a}{We assumed a 1~kpc distance for the blackbody emission radius at infinity.}
\tablefoottext{b}{The absorbed model flux covers the 0.2--10~keV range.}
\tablefoottext{c}{The abundance is fixed to the solar value.}
\tablefoottext{d}{We assume a canonical neutron star with 1.4~M$_\odot$ and 12~km radius.}
}
\end{table*}


\begin{figure}[t]
\begin{center}
\includegraphics[width=\linewidth]{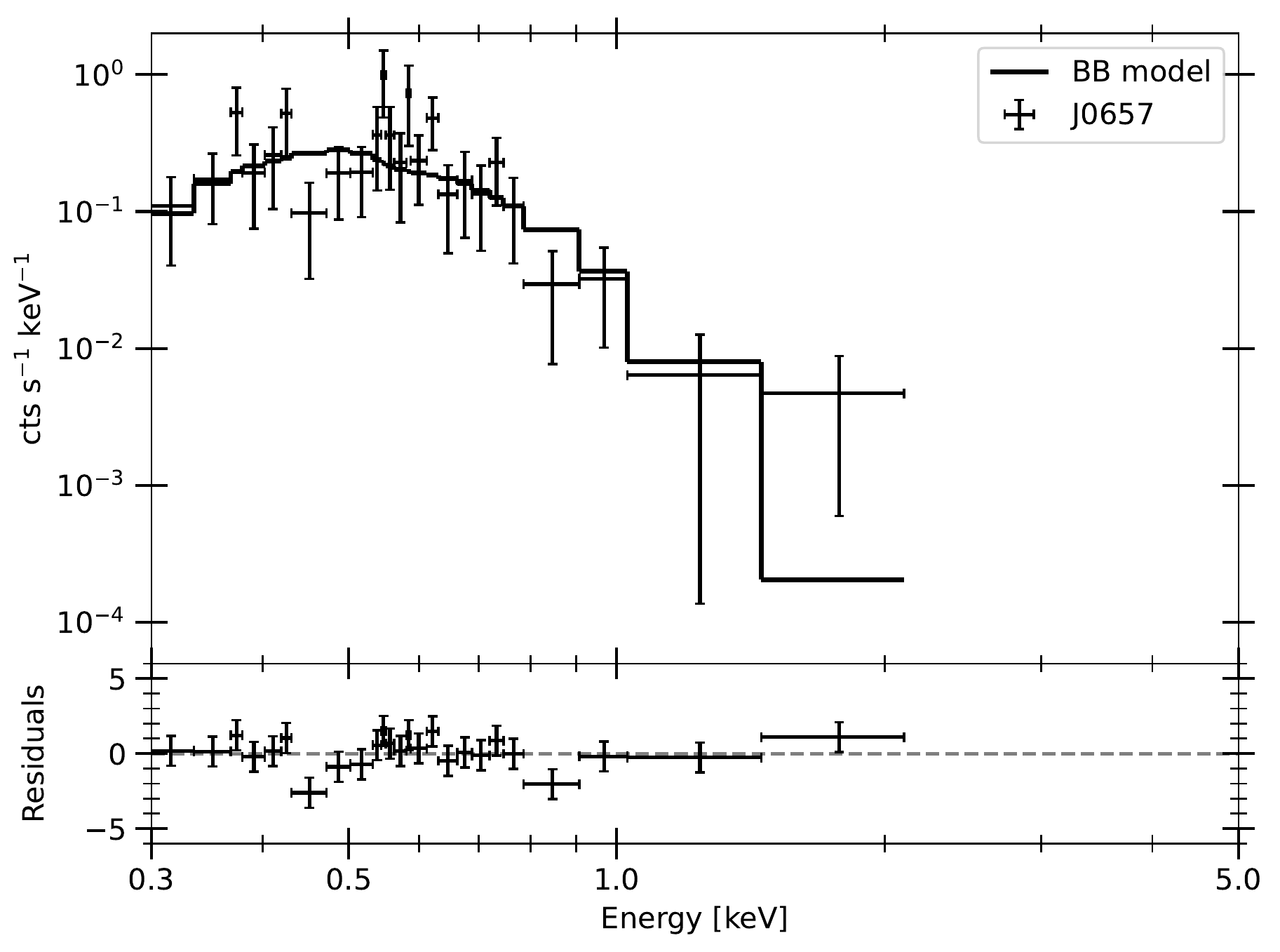}\vskip1pt
\includegraphics[width=\linewidth]{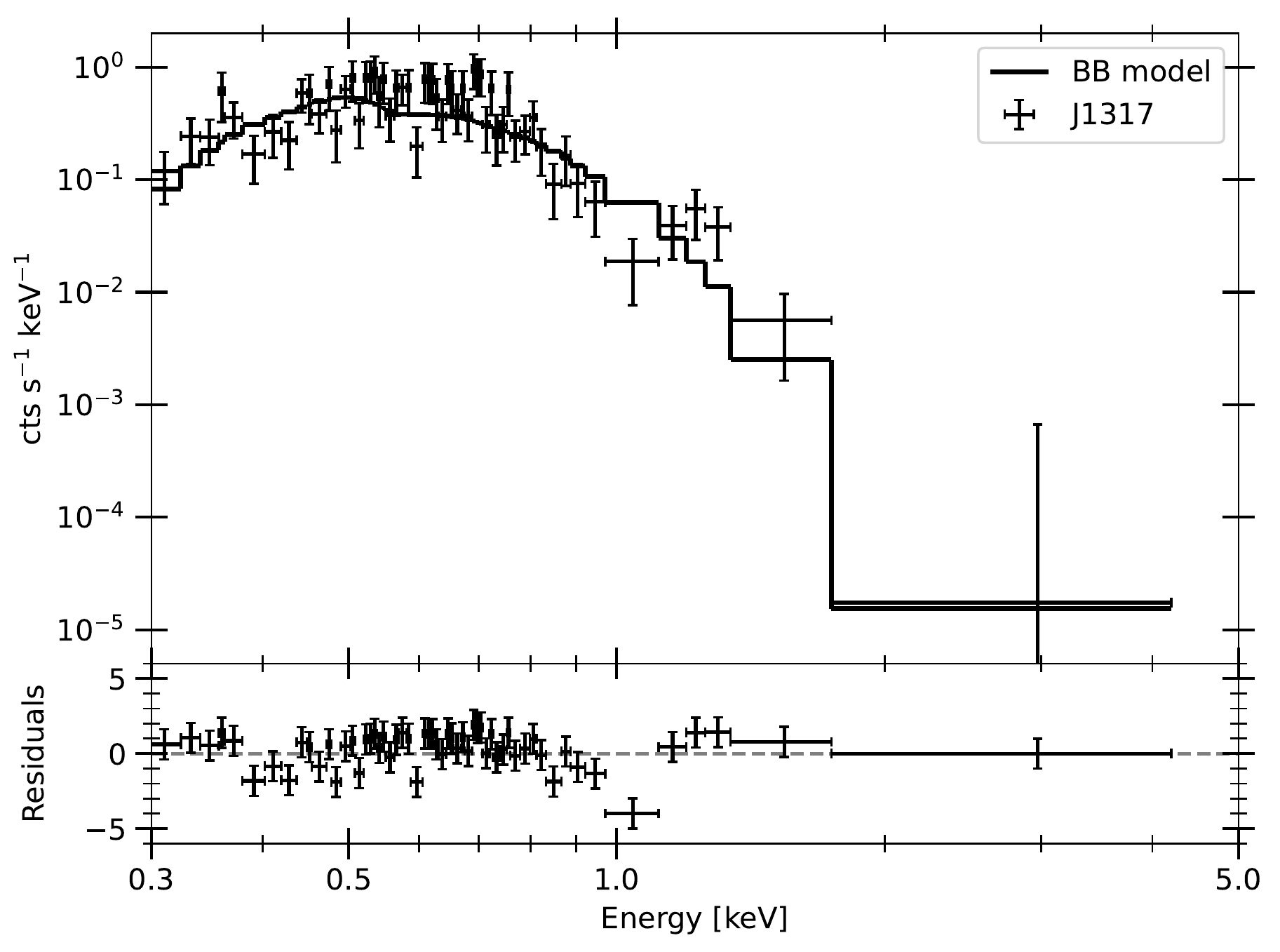}
\end{center}
\caption{Stacked eRASS spectra for \zsfs\ (top) and \otos\ (bottom) along with the best-fit blackbody model (see Table~\ref{tab_specfit}). The spectra were rebinned for presentation purposes.}
\label{fig_spec_plots}
\end{figure}


\subsection{eRASS observations}

The eRASS observations are carried out with \eros\ \citep{2021A&A...647A...1P}, the primary instrument on board the {\it SRG} mission. \eros\ consists of seven Wolter-I X-ray telescopes (dubbed TM1, TM2,..., TM7; together referred to as the virtual telescope TM0). All telescopes are operated with a polyimide foil (the standard FILTER setup) to reduce contamination, while an aluminium filter is applied directly on the CCDs for TM 1, 2, 3, 4, and 6 (together dubbed virtual telescope TM8). For TM5 and 7 (dubbed TM9), only the polyimide foil contains a thin aluminium layer, making them more sensitive to soft X-rays. However, this configuration leads to a time-variable optical light leak \citep{2021A&A...647A...1P}, which negatively affects the low energy calibration and performance. We list the available \eros\ observations for both candidates in Table~\ref{sec_obs}.

To enable an analysis beyond the catalogue properties, we adopted the photon event lists produced with the most recent pipeline processing version c020 and analysed the data with the eSASS software package \citep[version: eSASSusers\_211214.0.4.2, ][]{2022A&A...661A...1B}. We filtered the observations for periods of low background with the \texttt{flaregti} task. The total net exposures are 598~s and 1397~s of 'good' time (single visit times are presented in Table~\ref{sec_obs}) for \zsfs\ and \otos, respectively. We accepted all valid photon patterns (PATTERN$\leq$15) and excluded flagged or corrupted events, that is, those coming from bad pixels by applying FLAG=0xC000F000. We conducted a source detection run in the 0.2--10~keV band, as described in \cite{2022A&A...661A...1B}, with the goal of locating the target and nearby field sources via a maximum likelihood method of point spread function fitting. The output was used to create optimised source and background regions (Fig.~\ref{fig_reg}) that we applied to generate the spectra via the \texttt{srctool} task. We then grouped the spectra with one count per spectral bin using the \texttt{\mbox{grppha}} \citep[version: 3.1.0,][]{1995ASPC...77..367B} task. The source detection results are in agreement with the catalogue values listed in Table~\ref{tab_erass_prop}. Nevertheless, the catalogue offers the advantage that systematic errors can be better taken into account. Namely, by comparing the full eRASS:4 catalogue to the 'astrometric selection' from the \gaia\ DR3 qso\_candidate catalogue \citep{2022arXiv220605681G}, a global systematic error of 0.6\arcsec\ and an error scaling factor of 1.1 can be computed (see caption of Table~\ref{tab_erass_prop}). We present the single counts per visit in Table~\ref{sec_obs}. Combining all telescopes and observations, 96 counts and 418 counts (0.2--5~keV band) were detected for \zsfs\ and \otos.

\subsection{Optical follow-up observations}

We obtained deep optical imaging of \zsfs\ in December 2021 and March 2022 (Table~\ref{sec_obs}), using the Multi-Object Double Spectrograph 1 and 2 \citep[MODS,][]{2010SPIE.7735E..0AP} at \lbt. Each MODS splits the incoming light into a blue and red channel, allowing the simultaneous observation in the $g$- and $r$-band. In each visit we obtained 4~$\times$~50~s exposures with each MODS, leading to a total exposure time of 400~s in the $g$- and $r$-band. Unfortunately, the visit in March 2022 was affected by cirrus clouds, substantially increasing the sky background. For the image reduction, we used the most recent calibration images available for the respective visit. We first subtracted the bias from the science, dome, and sky flat frames and then divided the science images by a combined flat image created from the dome and sky flats.
 
The field of the INS candidate \otos\ was observed with \salt\ in May 2022 (Table~\ref{sec_obs}), using the Robert Stobie Spectrograph \citep[RSS,][]{2003SPIE.4841.1463B, 2003SPIE.4841.1634K} equipped with a fused silica clear filter. We obtained 40~$\times$~60~s exposures in imaging mode and adopted the pipeline-processed images (version: 0.42). The pipeline corrects for gain and cross-talk, as well as the bias estimated from the overscan region; it also mosaics the CCDs. In the absence of a proper flat field correction, the images were divided by a median normalised background image, created using the \texttt{photutils} package \citep{2022zndo...6385735B}. 

For both targets we derived the astrometric solution via the \texttt{astrometry.net} \citep{2010AJ....139.1782L} code and aligned the single images via the \texttt{wcsalign} \citep{astropy:2013, astropy:2018, astropy:2022} method. The images were then added to reach the full depth.

\section{Results\label{sec_analysis}}
\subsection{X-ray data analysis}
\subsubsection{Methodology}

We adopted the Bayesian X-ray Analysis \citep[BXA,][]{2014A&A...564A.125B} package in combination with XSPEC to fit models to the 'stacked' (which combine events from all eRASS scans) TM0 spectra of both candidates. To account for the interstellar absorption in the direction of the X-ray source, we applied the \texttt{tbabs} model component and elemental abundances of \cite{2000ApJ...542..914W}. For the continuum, we tested a simple blackbody model (\texttt{BB}), a power law (\texttt{PL}), the emission spectrum of a hot, collisionally ionised, plasma (\texttt{apec}\footnote{http://atomdb.org/}), and the neutron star atmosphere (\texttt{NSA}) model \citep[\texttt{}][]{1995ASIC..450...71P, 1996A&A...315..141Z}. We left most parameters free to vary and fixed only the \texttt{apec} model abundance to the solar value. The parameter ranges were chosen to cover the physically acceptable range of parameter values. For parameters spanning multiple magnitudes (such as $\nh$ ranging from $10^{18}$-$10^{23}$~cm$^{-2}$), we used log-uniform priors, whereas otherwise uniform priors were utilised.

We present the fit results for the different continuum model cases in Table~\ref{tab_specfit}. To compare the single fits, we also list the Bayesian evidence ($\log(Z)$), as well as Bayes factors (log(Z$_\mathrm{BB}$/Z)) with respect to the \texttt{BB} result, and Akaike information criterion (AIC) values. The Bayesian evidence is also referred to as the model evidence and can be regarded as the probability that the given data can be produced by the model. The AIC describes the information loss between the process that created the observed data and the proposed model that is supposed to describe this process. In general, the model with the larger Bayesian evidence or smaller AIC value is to be preferred. 

The spectra of the two INS candidates, folded with the best-fit blackbody models, are shown in Fig.~\ref{fig_spec_plots}. To estimate the overall fit quality, we present quantile-quantile and corner plots, displaying the parameter correlations and sample point distributions, in Fig.~\ref{fig_fit_imgs_j0657} and \ref{fig_fit_diagplots_j1317}. The fit results do not significantly change if the TM8 or TM9 spectra are fitted instead. We conclude that the remaining calibration uncertainties of TM5 and TM7 do not significantly affect the estimated parameter values. This implies that statistical uncertainties dominate the spectra.


\subsubsection{\zsfs\label{sec_zsfs}}

\begin{figure*}[t]
\hspace*{2mm}
\begin{minipage}{0.27\textwidth}
\begin{center}
\texttt{ tbabs~$\times$~bbodyrad}\\
\includegraphics[width=\linewidth]{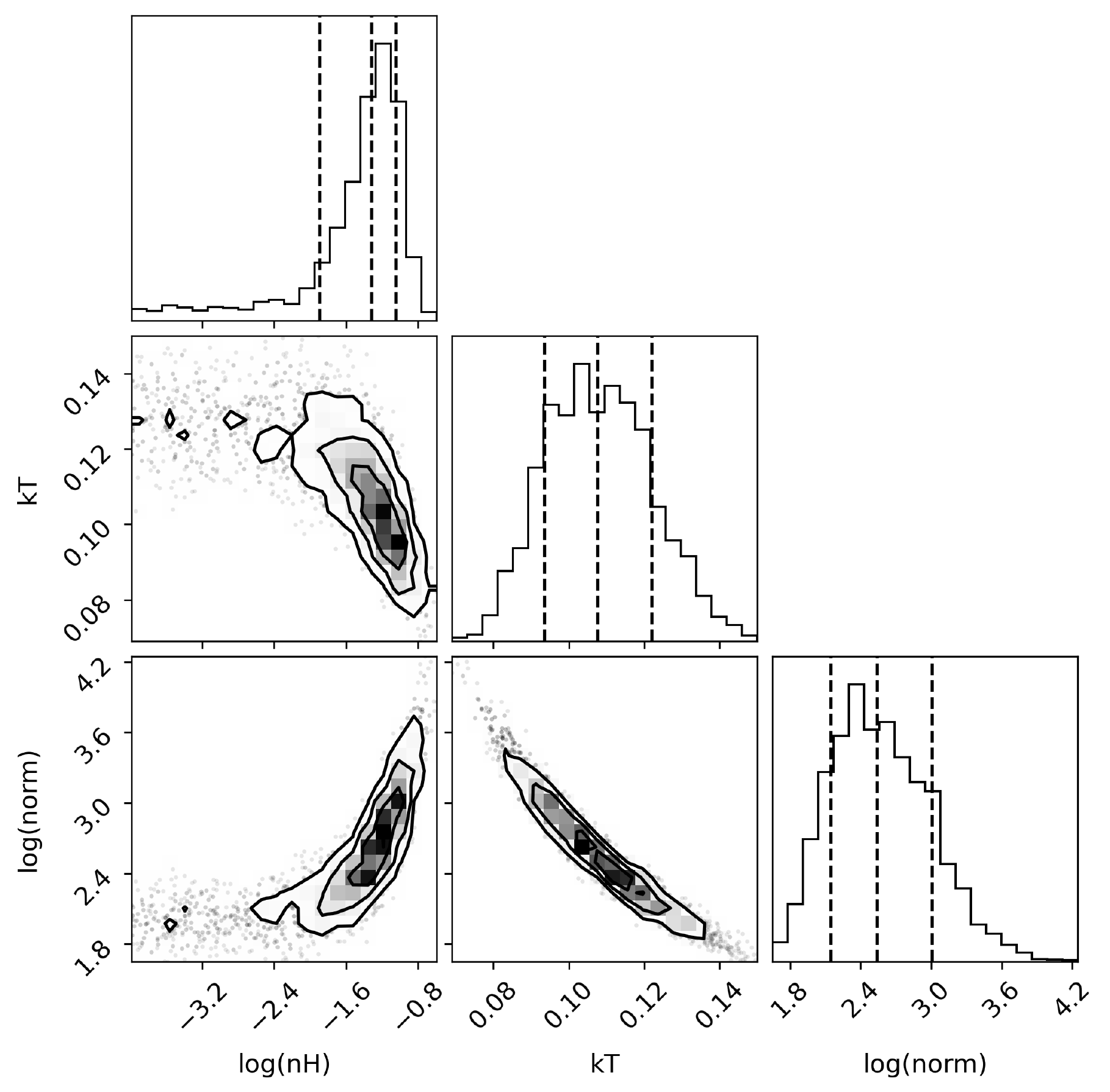}\vskip1pt
\includegraphics[width=\linewidth]{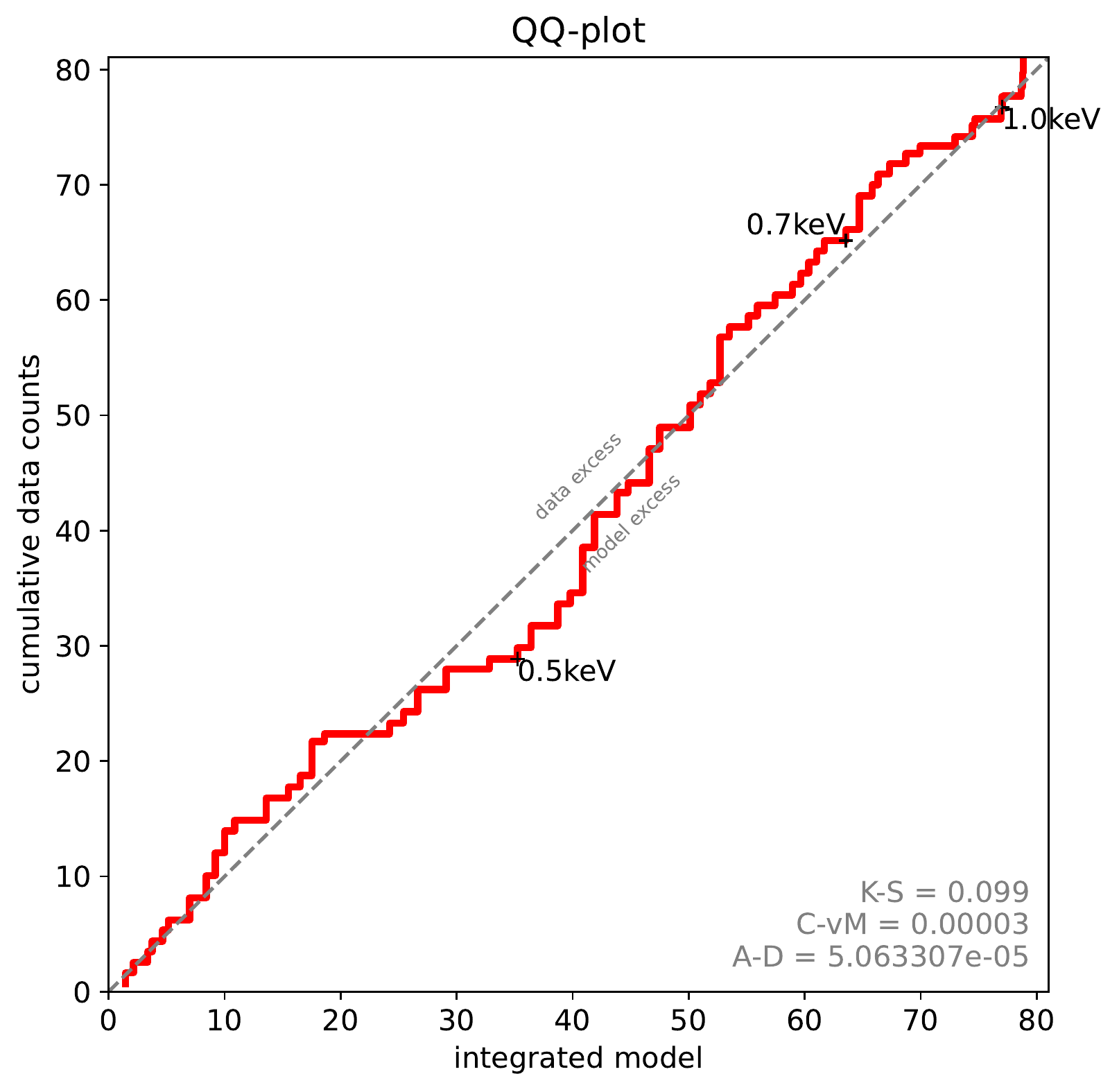}\vskip1pt
\end{center}
\end{minipage}\hspace*{14mm}
\begin{minipage}{0.27\textwidth}
\begin{center}
\texttt{ tbabs~$\times$~apec}\\
\includegraphics[width=\linewidth]{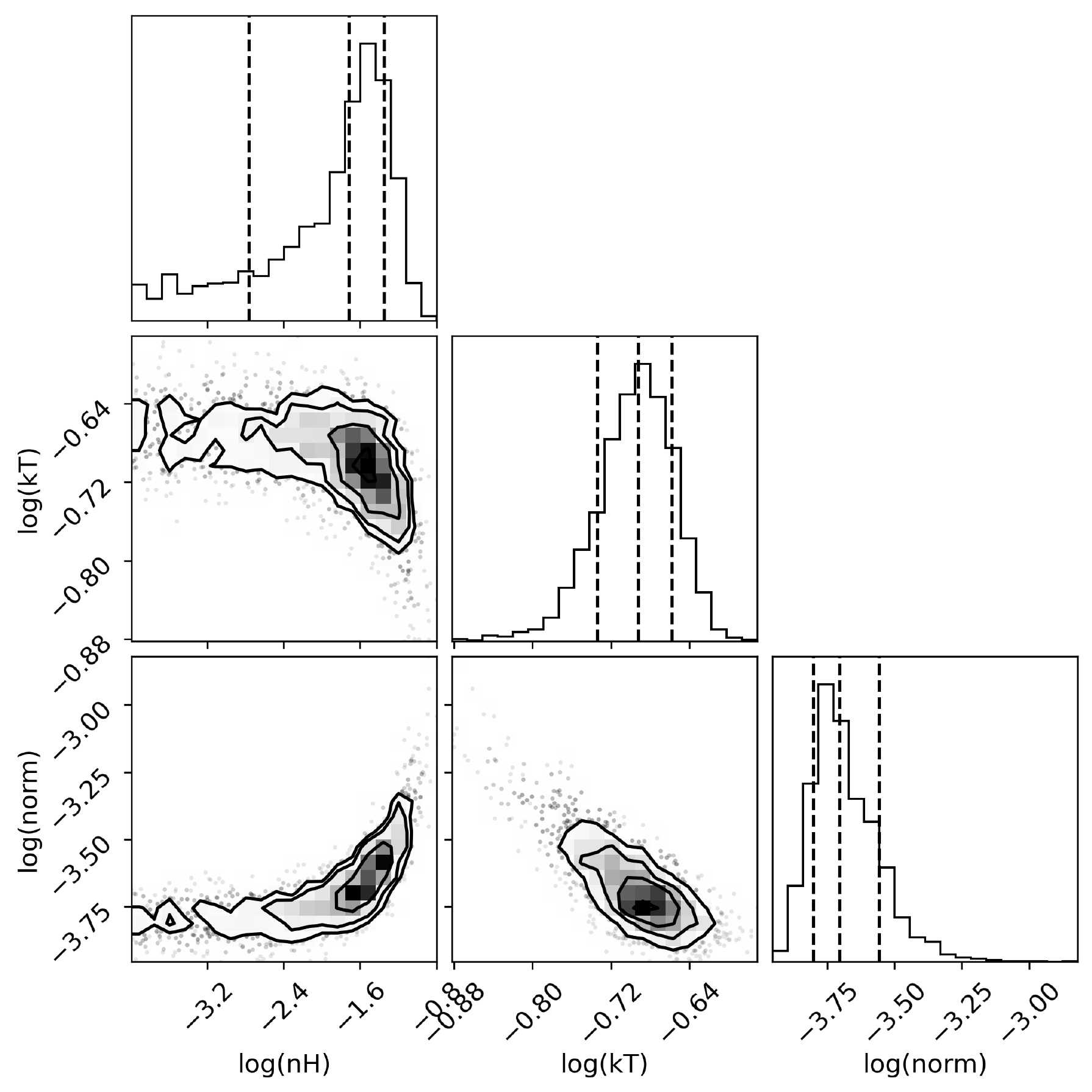}\vskip1pt
\includegraphics[width=\linewidth]{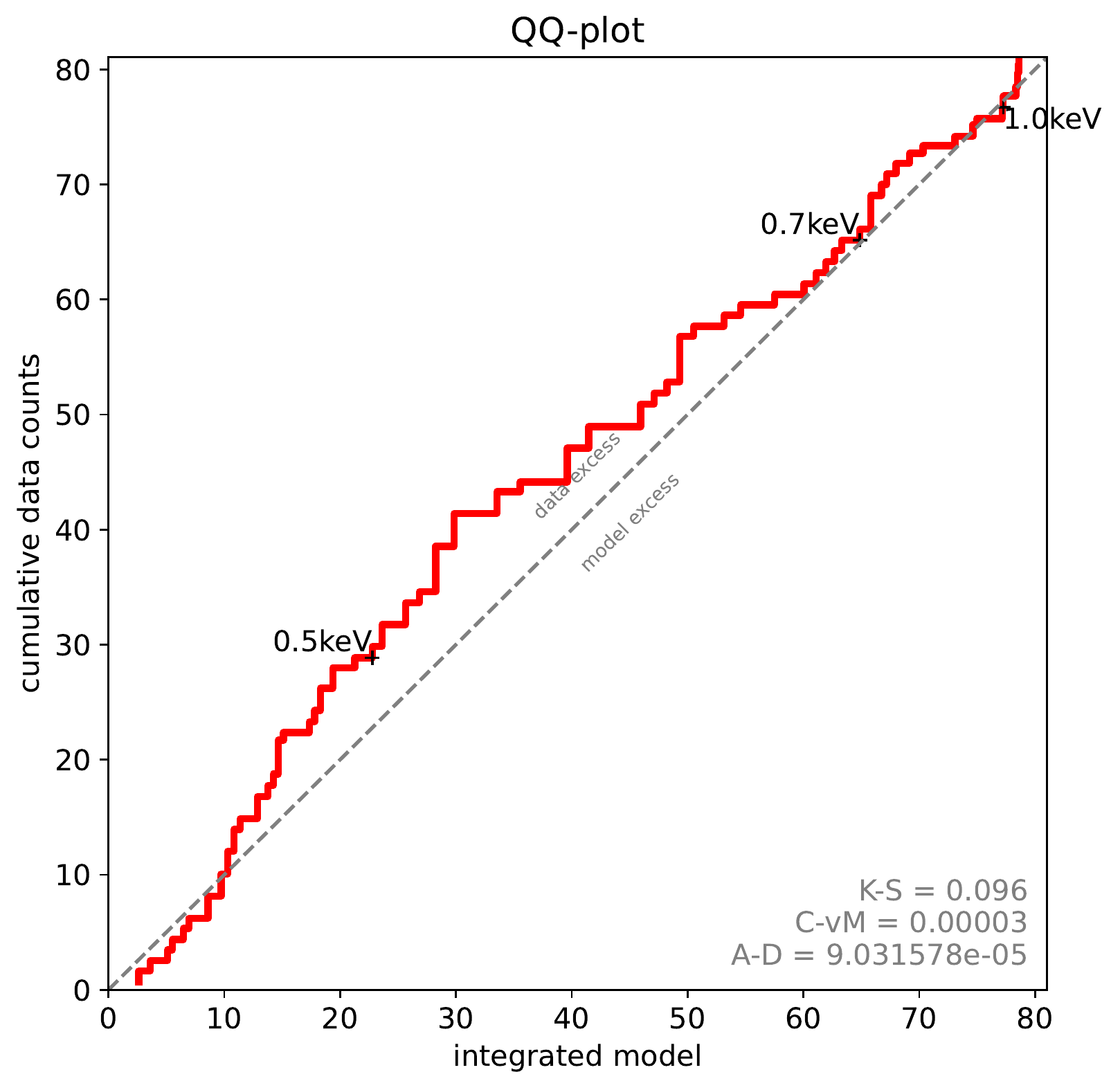}\vskip1pt
\end{center}
\end{minipage}\hspace*{14mm}
\begin{minipage}{0.27\textwidth}
\begin{center}
\texttt{ tbabs~$\times$~powerlaw}\\
\includegraphics[width=\linewidth]{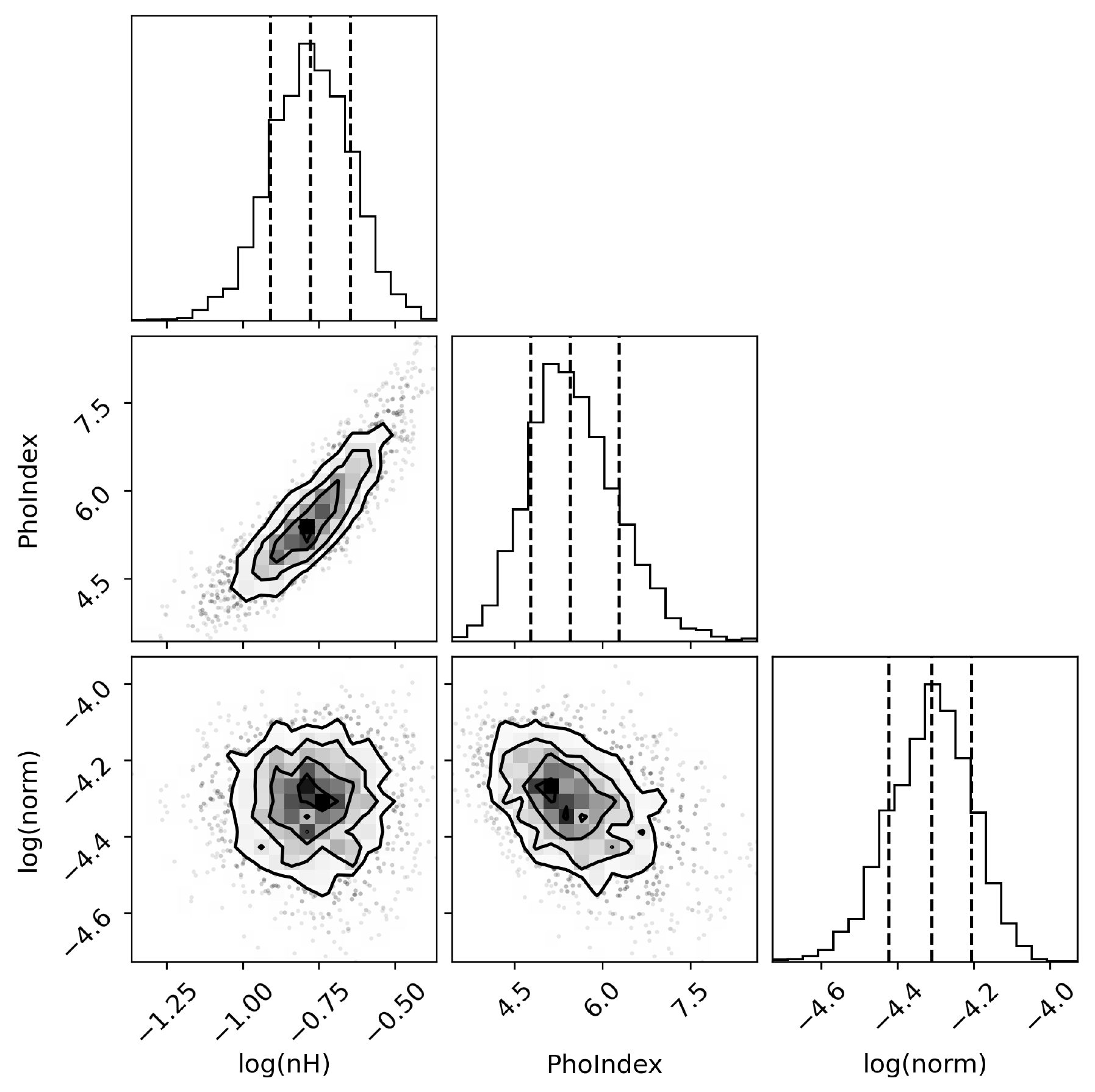}\vskip1pt
\includegraphics[width=\linewidth]{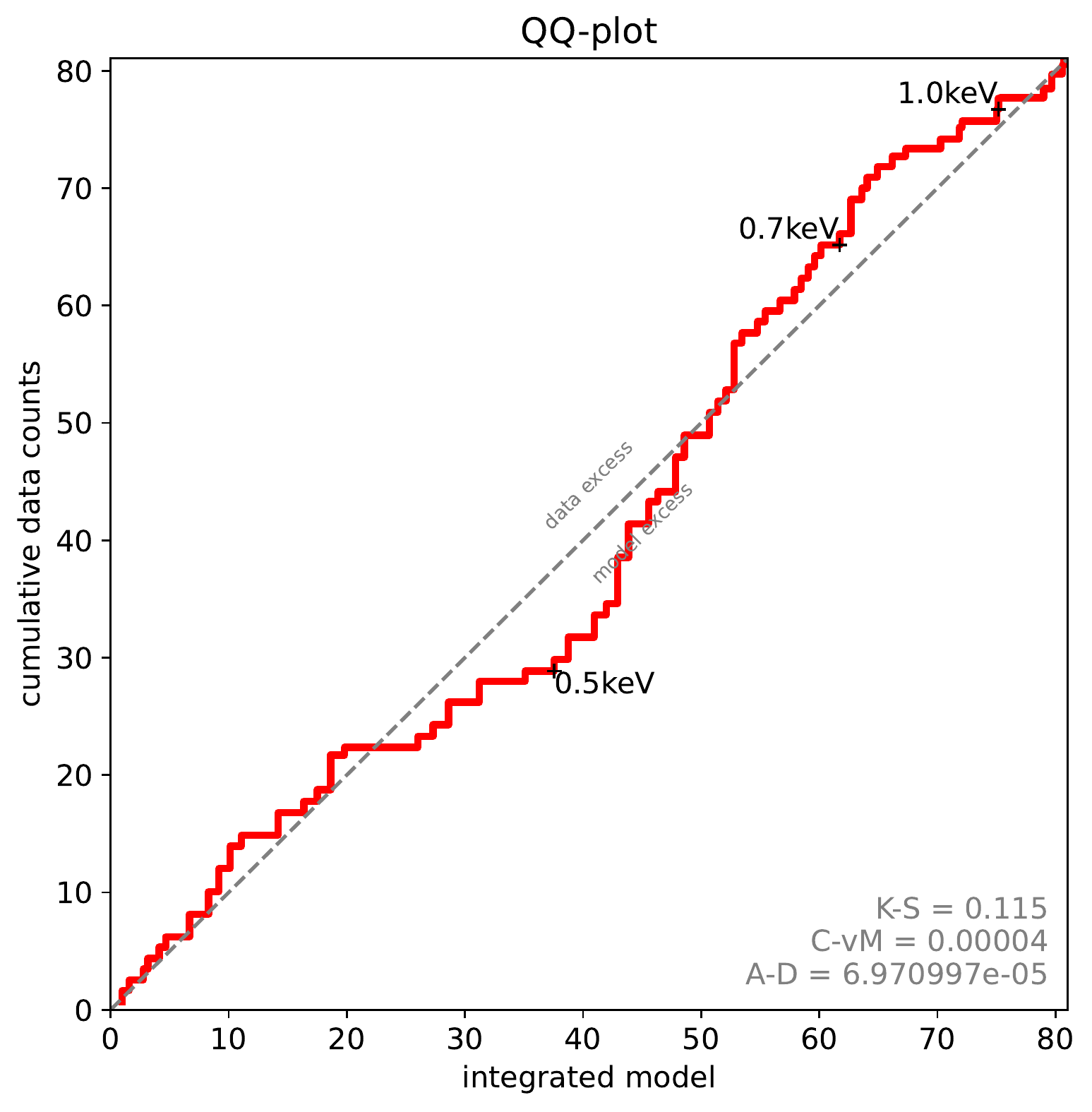}\vskip1pt
\end{center}
\end{minipage}
\caption{Diagnostic diagrams presenting the best-fit results for \zsfs\ for absorbed blackbody, apec, and power law spectra. For
each of the three models, the top row depicts corner plots presenting the parameter distribution and correlations based on the inferred sample of test points from the fit. The vertical dashed lines indicate the 16\%, median, and 84\% percentiles. The lower row contains quantile-quantile plots, comparing the cumulative sum of counts derived from the best-fitting model in comparison to the observed spectrum.}
\label{fig_fit_imgs_j0657}
\end{figure*}

\begin{figure*}[t]
\hspace*{2mm}
\begin{minipage}{0.27\textwidth}
\begin{center}
\texttt{tbabs~$\times$~bbodyrad}\\
\includegraphics[width=\linewidth]{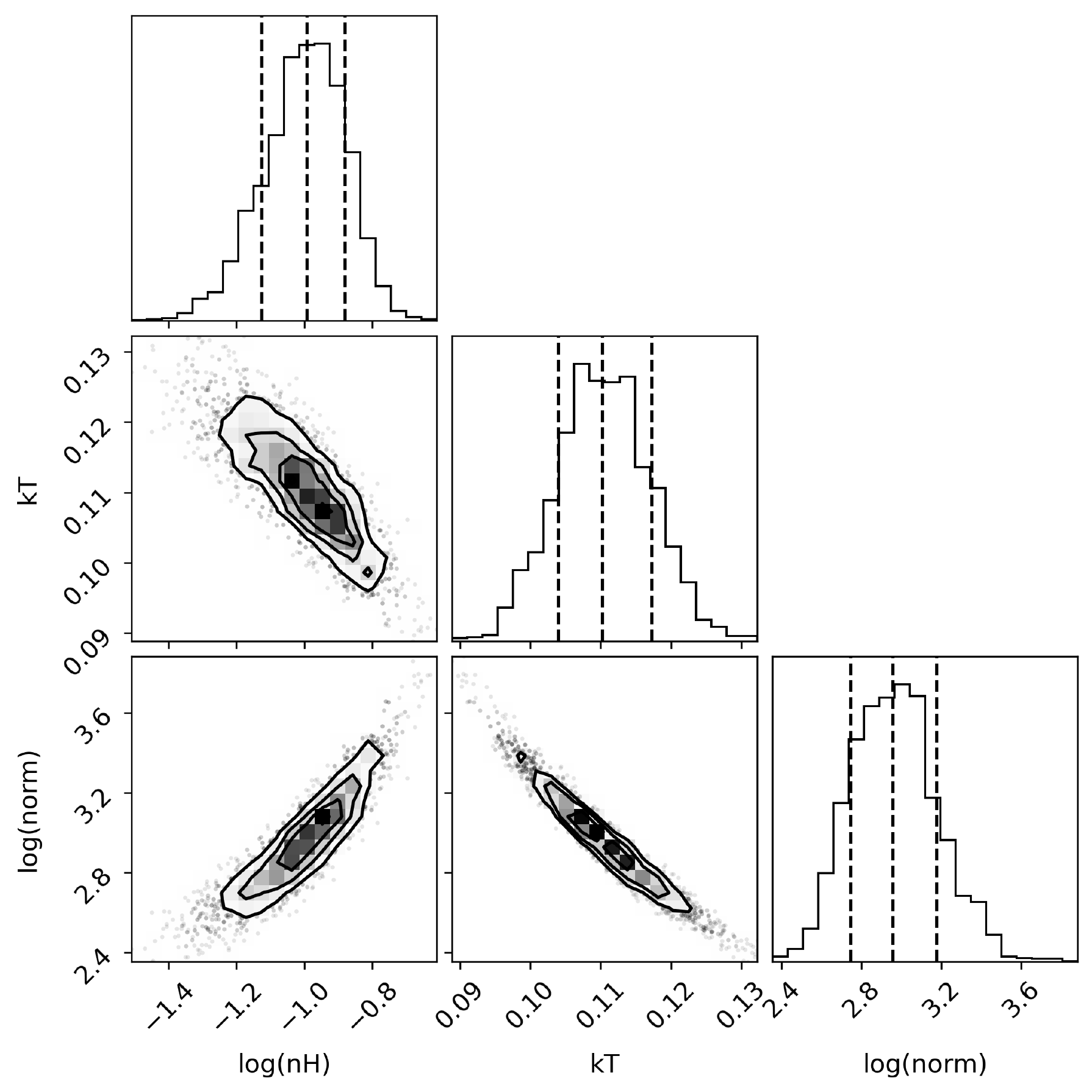}\vskip1pt
\includegraphics[width=\linewidth]{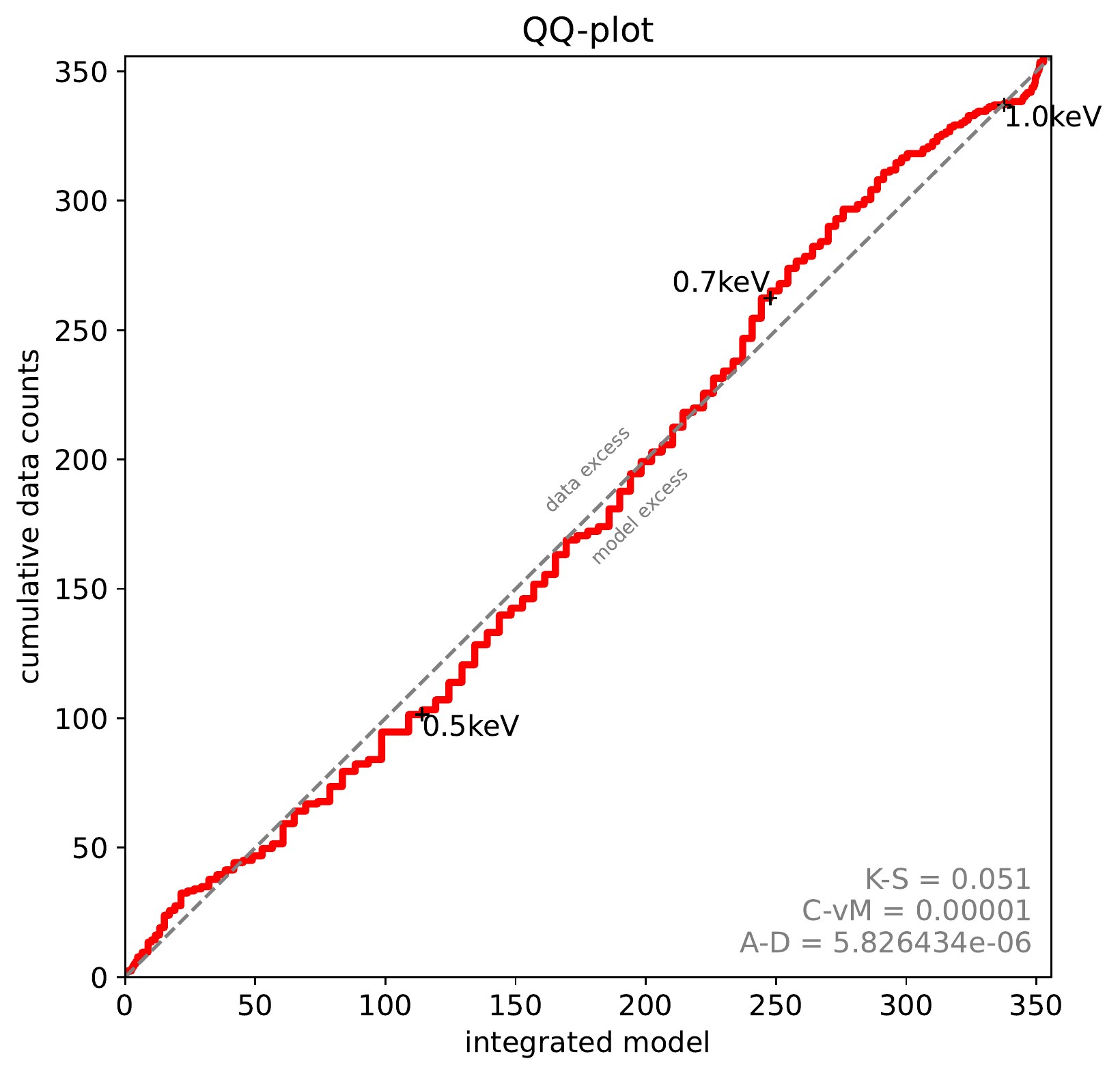}\vskip1pt
\end{center}
\end{minipage}\hspace*{14mm}
\begin{minipage}{0.27\textwidth}
\begin{center}
\texttt{tbabs~$\times$~apec}\\
\includegraphics[width=\linewidth]{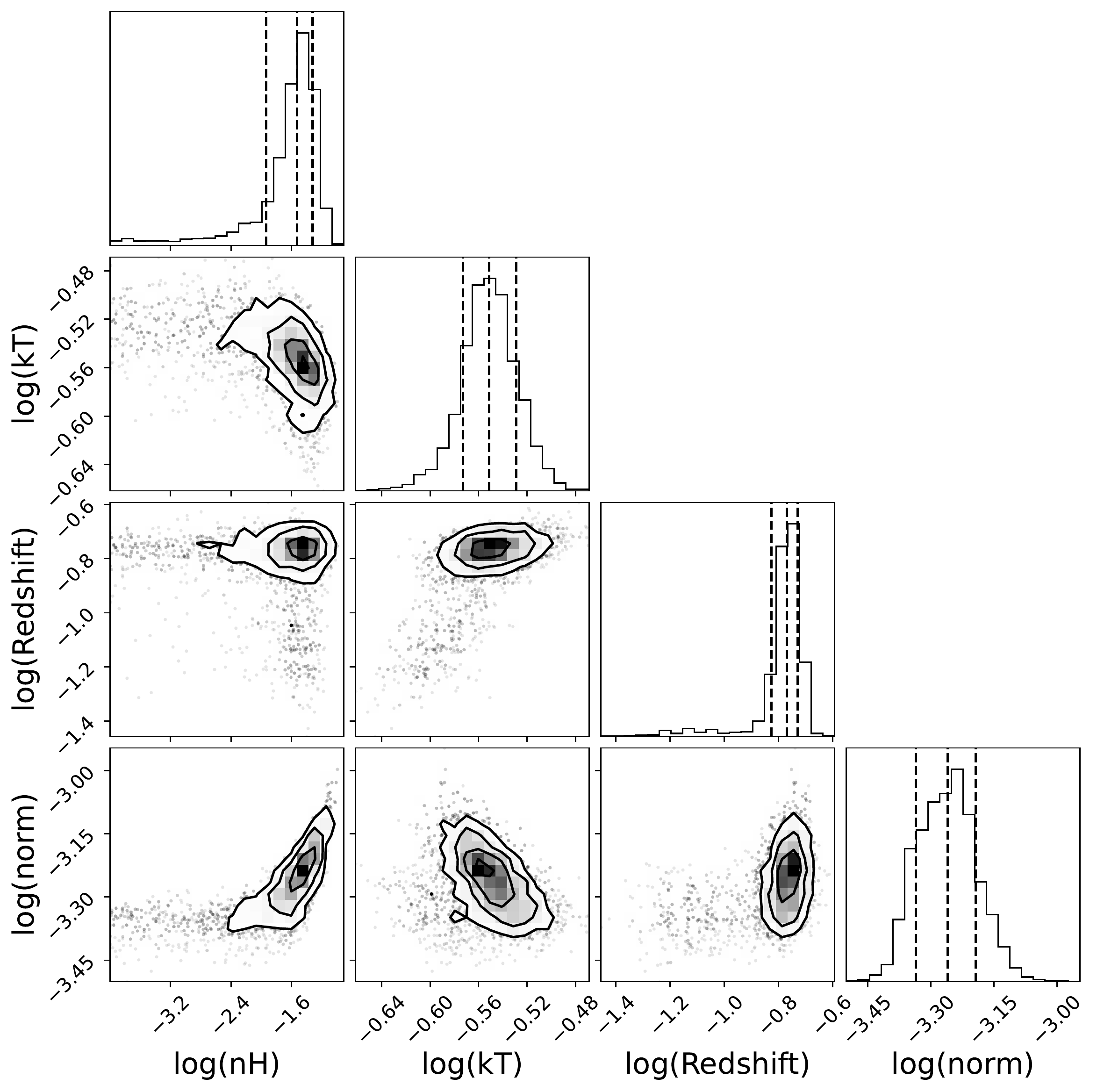}\vskip1pt
\includegraphics[width=\linewidth]{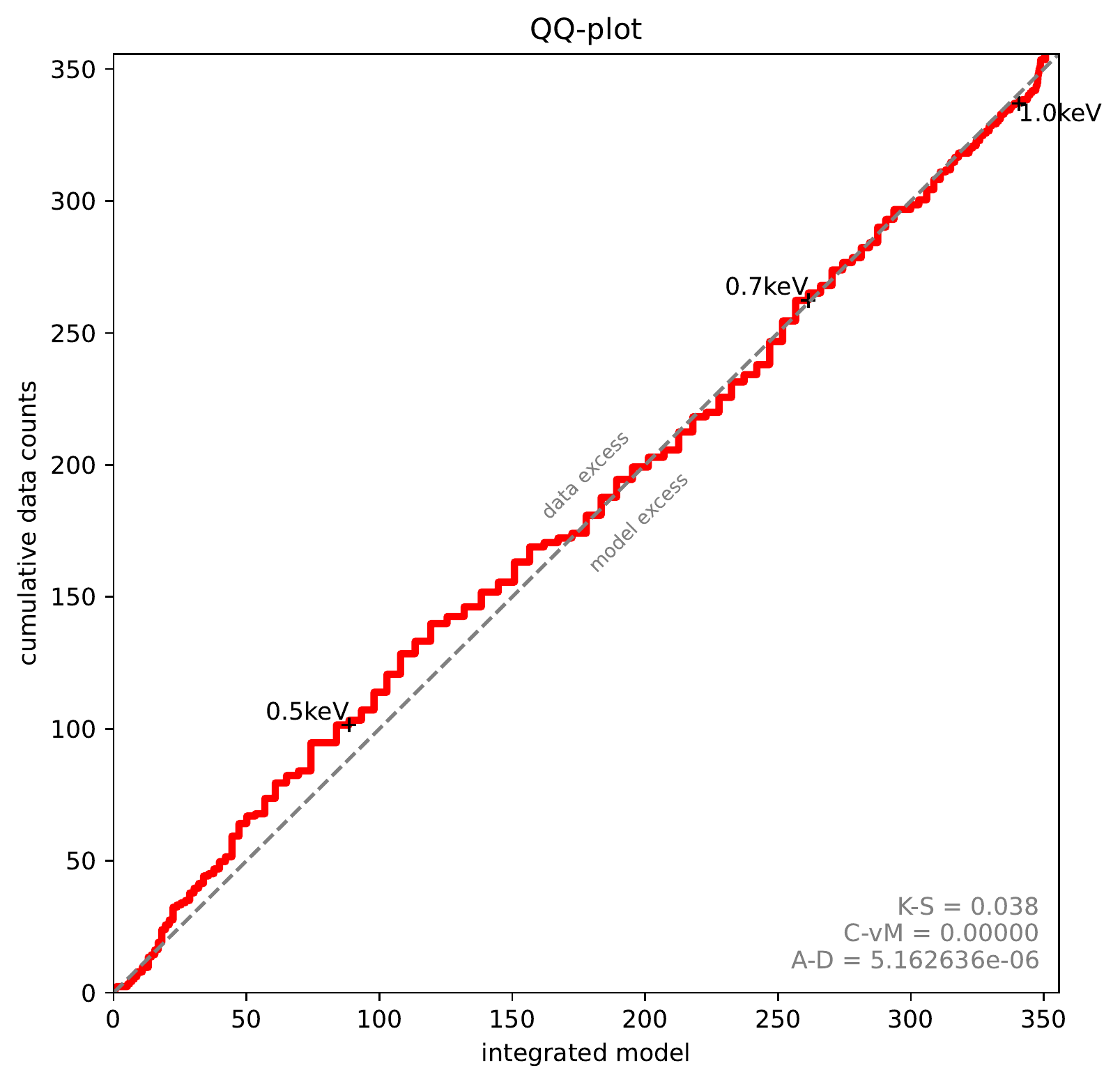}\vskip1pt
\end{center}
\end{minipage}\hspace*{14mm}
\begin{minipage}{0.27\textwidth}
\begin{center}
\texttt{tbabs~$\times$~powerlaw}\\
\includegraphics[width=\linewidth]{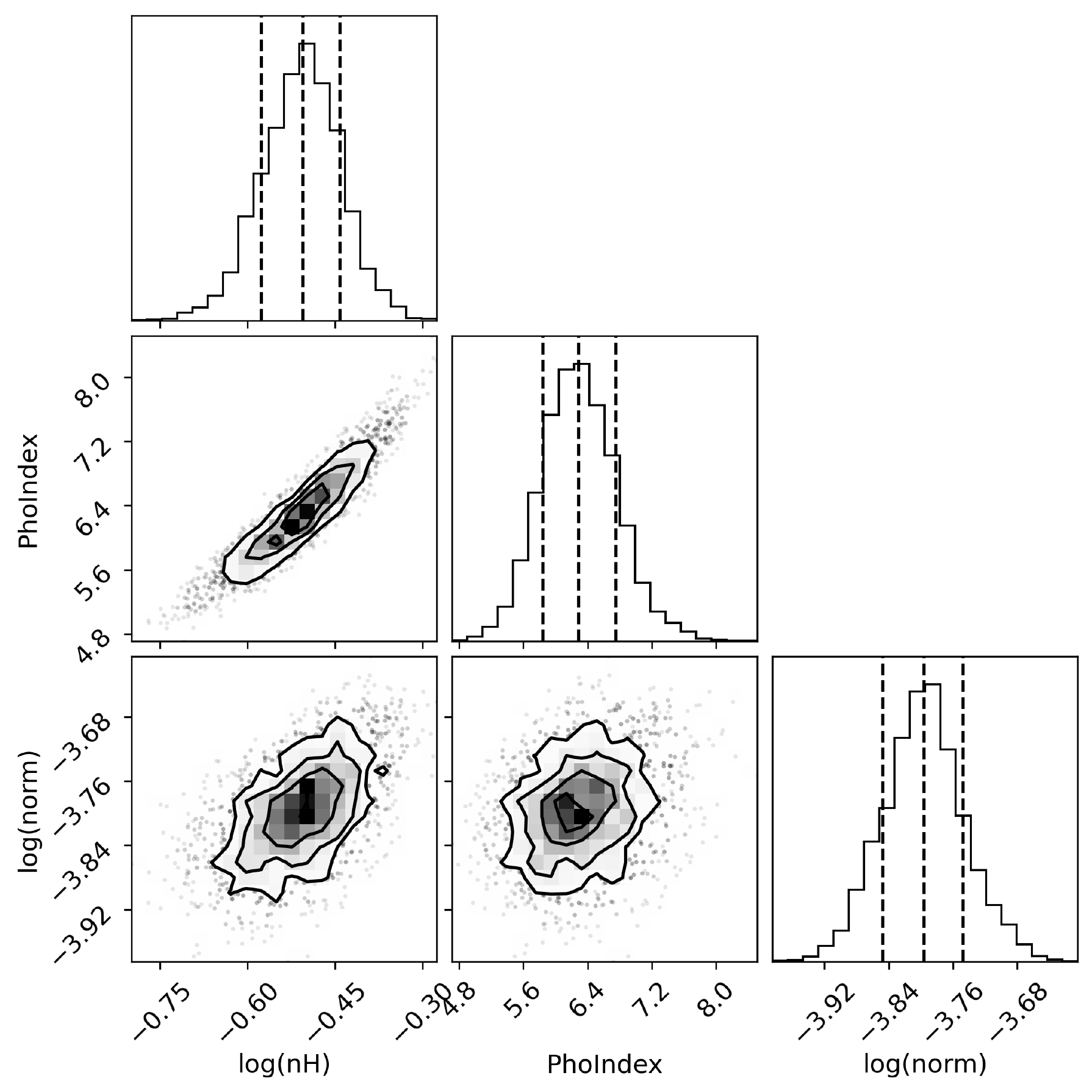}\vskip1pt
\includegraphics[width=\linewidth]{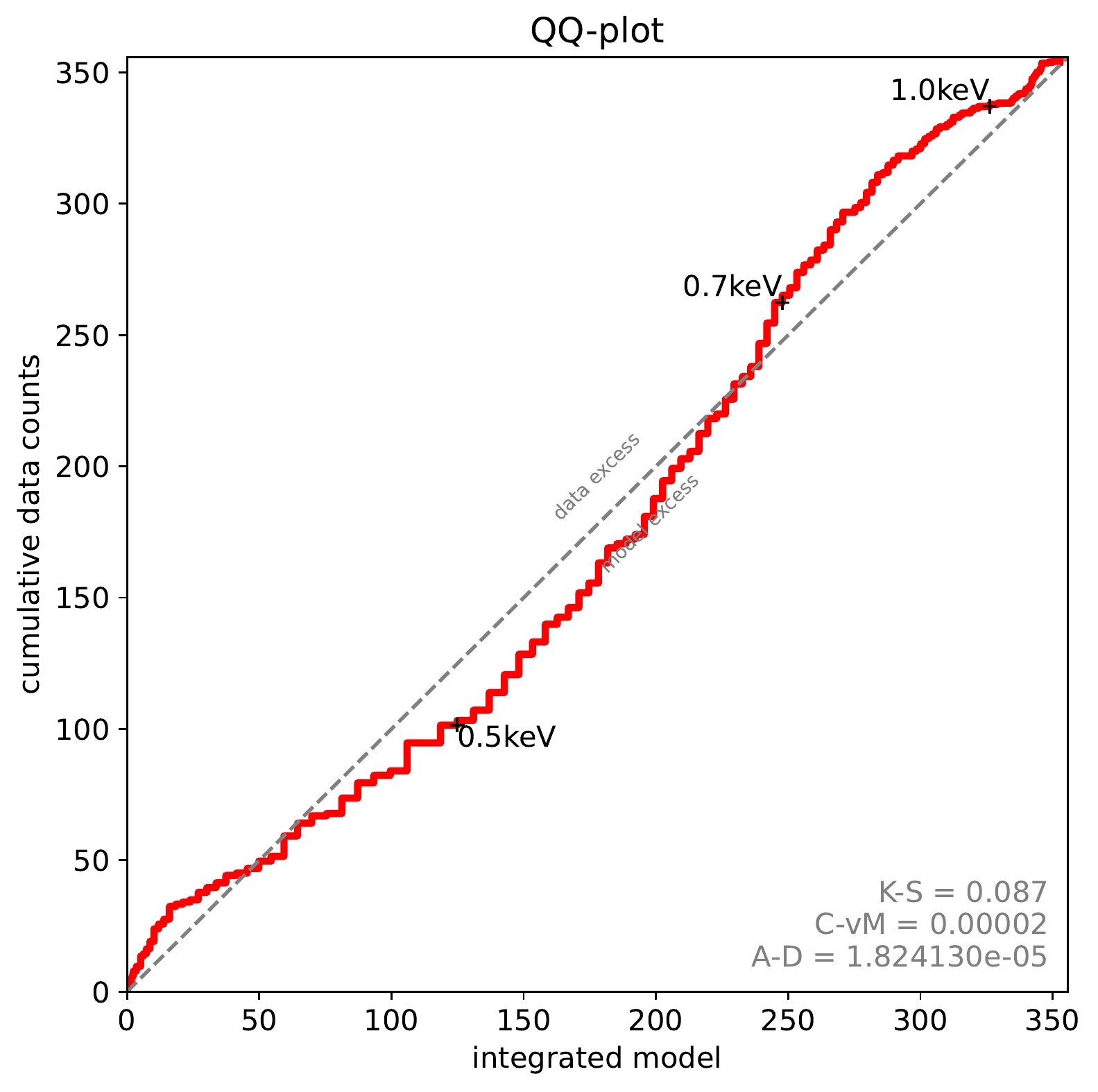}\vskip1pt
\end{center}
\end{minipage}
\caption{Diagnostic diagrams presenting the best-fit results for \otos\ for absorbed blackbody, apec, and power law spectra.  For
each of the three models, the top row depicts corner plots presenting the parameter distribution and correlations based on the inferred sample of test points from the fit. The vertical dashed lines indicate the 16\%, median, and 84\% percentiles. The lower row contains quantile-quantile plots, comparing the cumulative sum of counts derived from the best-fitting model in comparison to the observed spectrum.}
\label{fig_fit_diagplots_j1317}
\end{figure*}

The Bayes factors and AIC values presented in Table~\ref{tab_specfit} favour a \texttt{BB} model with $kT= 108^{+15}_{-14}$~eV. Assuming a 1~kpc distance to the source, we find a $1.9^{+1.4}_{-0.7}$~km emission region radius. A \texttt{PL} fit results in worse log(Z) and AIC values and converges to a photon index of $6.5^{+0.9}_{-0.7}$, which is unreasonably steep for most AGN types \citep[usual photon indices of 1.5--2.5, ][]{2010A&A...512A..58I}, and $\nh$ values exceeding the Galactic column density (presented in Table~\ref{tab_erass_prop}), which could support an extragalactic nature. We find the redshift parameter of the \texttt{apec} model to be consistent with zero and it was thence fixed. The fit resulted in a temperature $\sim 200$~eV and in $\nh$ values below the Galactic column density, both in accordance with a stellar nature. The AIC and Bayesian evidence are, in comparison to the \texttt{BB} fit, still worse and we note that the quantile-quantile plot (Fig.~\ref{fig_fit_imgs_j0657}) shows \texttt{apec} to systematically underpredict the source emission below energies of 0.7~keV.

We used the best-fit solutions of Table~\ref{tab_specfit} to generate 200 \texttt{BB}, \texttt{PL}, and \texttt{apec} spectra and fitted the data with all three model types. The goal of the simulations is to evaluate the false-positive (the best-fit model is not the true model) and false-negative (the true model is rejected) rates, on the basis of the Bayes factors. The simulations show that the \texttt{apec} model is erroneously accepted in 1.5\% of the cases for \zsfs, assuming that it emits as a blackbody. On the other hand, the small ($1.6$\%) false-positive error of the blackbody disfavours \texttt{apec} as the underlying source spectrum. In the case of a \texttt{BB} nature, we never found a \texttt{PL} model to fit the spectrum better. Instead, we found a large false-negative rate for the steep \texttt{PL} spectrum, since a \texttt{BB} nature was favoured over the true \texttt{PL} in 65\% of the cases. Dedicated, deeper observations, with better count statistics, are needed to discern between a \texttt{BB} and a \texttt{PL} model. As it can be seen for the case of \otos\ (see Sect~\ref{sec_otos}), a better exposed source spectrum can drastically reduce the false-negative rate and thereby also lead to generally worse \texttt{PL} fits.

We present the results of \texttt{NSA} fits in Table~\ref{tab_specfit}. We assumed a canonical NS with 12~km radius and a mass of 1.4~M$_\odot$. The fit quality is between those of the \texttt{BB} and \texttt{apec} fits; however, the \texttt{NSA} result converges to a 20--50\% lower temperature, while the column density $\nh$ exceeds the Galactic value (see Table~\ref{tab_erass_prop}). The difference in AIC and log(Z) are not sufficient to favour a particular magnetic field strength. The inferred distances are well in agreement with those of the known XDINS population. 

We found no model combination (for example a \texttt{BB+PL} model that also accounts for non-thermal emission) that significantly improves the results of the single component models presented in Table~\ref{tab_specfit}. Nevertheless, one can estimate an upper limit for the non-thermal to thermal flux ratio, by fitting a \texttt{BB+PL} model with the \texttt{PL} photon index fixed to two. The fit converged to a \texttt{BB} component with $103^{+16}_{-14}$~eV, $\nh$ of $6^{+5}_{-4} \times 10^{20}$~cm$^{-2}$, and an emission radius of $2.1^{+1.6}_{-0.9}$~km (at 1~kpc distance). The log(Z$_\mathrm{BB}$/Z)=0.5 and AIC=65.3 indicate the fit to be of a similar quality to the single \texttt{BB} solution presented in Table~\ref{tab_specfit}. We computed $F_\mathrm{PL}/F_\mathrm{BB} \leq 56\%$ from the upper and lower $3\sigma$ power law and blackbody component flux limits located at $1.8 \times 10^{-13}$~\fluxcgs\ and $3.2\times 10^{-13}$~\fluxcgs\ (0.2-10~keV). 

\subsubsection{\otos\label{sec_otos}}

We found a \texttt{BB} with a temperature of $110^{+7}_{-7}$~eV to maximise the Bayesian evidence. The emission region radius at a 1~kpc distance computes to $3.0^{+0.9}_{-0.7}$~km. These values are in agreement with the expectation for an INS, but we found the fit to converge to a $\nh$ value exceeding the Galactic limit (see Table~\ref{tab_erass_prop}). This is at odds with a Galactic INS interpretation, but could also hint towards a more complicated spectral shape. Contrarily, the AIC favours an \texttt{apec} fit that converges to a temperature of $281^{+15}_{-14}$~eV, coupled with a redshift of $0.170^{+0.016}_{-0.022}$ that might indicate an extragalactic nature. Interestingly, here the $\nh$ converged to a result that is below the Galactic value and, as such, points towards the object being of Galactic origin. If one attempts to fix the redshift to zero, the fit quality worsens (log(Z$_\mathrm{BB}$/Z)=13.87, AIC=153.67). A \texttt{PL} fit results in a photon index of $7.3^{+0.5}_{-0.5}$, which is again too steep for most known AGN types. The resulting log(Z) and AIC are clearly worse in comparison to the \texttt{BB} and \texttt{apec} models.

As for \zsfs, we simulated 200 spectra, using the best-fit solutions of Table~\ref{tab_specfit} to estimate the false-positive and false-negative rates. Using the AIC as criterium, we find a large false-positive rate of 19\% indicating \texttt{apec} to fit better, although we assumed \texttt{BB} emission for the source. Interestingly, for the same fits, the resulting Bayes factors never favoured the \texttt{apec} model over a \texttt{BB} fit. We conclude that the observed AIC, indicating \texttt{apec} to fit best, is likely a spurious result, although larger differences in AIC, as observed here, were found only for 1.5\% of our simulations. On the other hand, the false-positive rate is very small (a true \texttt{apec} spectrum is never better described by a \texttt{BB} model). Thus, choosing the \texttt{BB} over the \texttt{apec} model seems to be robust.

We found no case where a \texttt{BB} spectrum was better fit by a \texttt{PL} than a \texttt{BB} model. Nevertheless, the false-positive rate of 8\% -- favouring a \texttt{BB} model, although the source has a \texttt{PL} nature -- is still high. This value is much lower than the one observed for \zsfs, which we attribute to the higher number of counts in the \otos\ case that allow for a better distinction between a \texttt{BB} and \texttt{PL} nature, but it is still difficult to reject an absorbed \texttt{PL} nature on the available fit alone.

The best-fit \texttt{NSA} model results, listed in Table~\ref{tab_specfit}, are generally worse than the \texttt{BB} and \texttt{apec} models. Comparing the Bayes factors, magnetised models seem to fit better and the distance estimates are in agreement with the known XDINS population. We again found the \texttt{NSA} model temperatures to be 25--50\% of the \texttt{BB} temperatures, while the $\nh$ is about twice the Galactic value, which is at odds with a Galactic nature of the source. 

Similarly to \zsfs, we found no multi-component models that significantly improve the fits in comparison to the single component models shown in Table~\ref{tab_specfit}. A \texttt{BB+PL} fit, with the photon index fixed to two, is worse in quality, but it converges to \texttt{BB} parameter values very similar to the single \texttt{BB} result (log(Z$_\mathrm{BB}$/Z)=1.0, AIC=129.5, $\nh = 10.7^{+4.0}_{-2.7} \times 10^{20}$~cm$^{-2}$, $kT=108^{+8}_{-8} $~eV). The fit results in an upper 3$\sigma$ \texttt{PL} flux limit of $1.3\times 10^{-13}$~\fluxcgs\ and a lower 3$\sigma$ \texttt{BB} flux limit of $1.1\times 10^{-12}$~\fluxcgs\ (0.2--10~keV). Thus, we found $F_\mathrm{PL}/F_\mathrm{BB} \leq 12\%$.


\begin{table}
\caption{Simultaneous blackbody fit results of the single eRASS spectra.
\label{tab_simspecfit}}
\centering
\begin{tabular}{ccccccccc}
\hline\hline
\multicolumn{4}{l}{\zsfs}\\
\hline
eRASS & $kT$ & Radius\,\tablefootmark{(a)} & Absorbed flux\,\tablefootmark{(b)}\\
& [eV] & [km] & [$10^{-13}$~\fluxcgs]\\
\hline
1 &  $85^{+16}_{-13}$ & $3.7^{+2.4}_{-1.5}$ & $2.5^{+0.8}_{-0.6}$\\ 
2 & $125^{+17}_{-18}$ & $1.3^{+0.7}_{-0.4}$ & $2.6^{+0.8}_{-0.7}$\\
3 &  $94^{+15}_{-12}$ & $3.1^{+1.7}_{-1.1}$ & $3.1^{+0.7}_{-0.7}$\\
4 & $119^{+19}_{-16}$ & $1.4^{+0.8}_{-0.5}$ & $2.0^{+0.6}_{-0.5}$\\
\hline
\multicolumn{4}{l}{\otos}\\
\hline
eRASS & $kT$ & Radius\tablefootmark{(a)} & Absorbed flux\tablefootmark{(b)}\\
& [eV] & [km] & [$10^{-13}$~\fluxcgs]\\
\hline
1 & $104^{+7}_{-7}$   & $3.7^{+0.9}_{-0.7}$ & $4.8^{+0.5}_{-0.5}$\\
2 & $113^{+9}_{-8}$   & $2.8^{+0.7}_{-0.6}$ & $4.4^{+0.6}_{-0.5}$\\
3 & $115^{+11}_{-9}$  & $2.8^{+0.8}_{-0.7}$ & $4.7^{+0.8}_{-0.7}$\\
4 & $110^{+10}_{-9}$  & $3.0^{+0.9}_{-0.7}$ & $4.3^{+0.6}_{-0.6}$\\
5 & $111^{+11}_{-10}$ & $2.6^{+0.9}_{-0.7}$ & $3.5^{+0.6}_{-0.6}$\\  
\hline

\end{tabular}
\tablefoot{We present the median value estimated from the sample parameter distribution and the errors give the 1$\sigma$ confidence region. We adopted 400 live points to sample the parameter space. The $\nh$ is fixed to the values estimated from the stacked spectra (see Table~\ref{tab_specfit}).
\tablefoottext{a}{We assumed a 1~kpc distance for the blackbody emission radius at infinity. }
\tablefoottext{b}{The absorbed model flux covers the 0.2--10~keV range.}
}
\end{table}


\subsection{Variability and timing}

The available \eros\ data, spanning multiple epochs (see Table~\ref{tab_obs}), allow one to check for flux variations. To this end, we conducted simultaneous fits to the single eRASS spectra with an absorbed blackbody component, fixing the $\nh$ to the value obtained from the fits to the stacked data, but leaving the normalisation and effective temperature free to vary between the different eRASS visits. The result is shown in Table~\ref{tab_simspecfit}. There are no hints towards a significant variation for either of the two candidates since the flux values are in agreement with each other within 1--2$\sigma$. Variations of this size were also observed for the known XDINSs and they can be attributed to low-count statistics (see Appendix B in \cite{2022A&A...666A.148P} for a more detailed discussion). For \otos, there is a source, 27\arcsec\ away from the eRASS position, in the second \ros\ all-sky survey source catalogue \citep{2016A&A...588A.103B}. Considering the $1 \sigma$ positional uncertainty of 20\arcsec, the identification seems robust, mostly since there are no similarly bright eRASS sources in a 5\arcmin\ radius around \otos. We used the sample point distribution of the best-fit \texttt{BB} model from Table~\ref{tab_specfit} to simulate the expected \ros\ count rate with XSPEC. We found a count rate of $2.9^{+1.5}_{-0.8} \times 10^{-2}$~s$^{-1}$ (0.1--2.4~keV) which is in agreement with the \ros\ catalogue count rate of $6.5(1.8) \times 10^{-2}$~s$^{-1}$ (0.1--2.4~keV) within 2$\sigma$. \zsfs,\ on the other hand, was not detected by ROSAT. From five ROSAT sources within 45\arcmin\ of the eROSITA position, we determined that ROSAT was pointed for about 430~s in the direction of \zsfs, which would result in an expected ROSAT count rate of $1.9^{+1.2}_{-1.0} \times 10^{-2}$~s$^{-1}$ (0.1--2.4~keV). This is at the detection limit, as can be seen from the lower panel in Fig.~8 of \cite{2016A&A...588A.103B}.

As the sources were observed in scanning mode, a search for pulsations is not feasible. For \zsfs, only about 17--25 photons (0.2--5.0~keV) are detected in each visit (Table~\ref{tab_obs}). The largest photon count (105 photons) was achieved for \otos\ during eRASS1. We applied a Z$^2_m$ periodicity search \citep{1983A&A...128..245B} and did not find any significant pulsations. For pulsations within $P=100$~ms -- $20$~s (using events within 0.2--2.0~keV and $1.3\times 10^6$ independent trials), this provides a non-constraining 3$\sigma$ upper limit for the pulsed fraction of 87\%.


\subsection{Optical imaging}
\begin{figure*}[t]
\begin{minipage}{0.49\textwidth}
\begin{center}
\includegraphics[width=\linewidth]{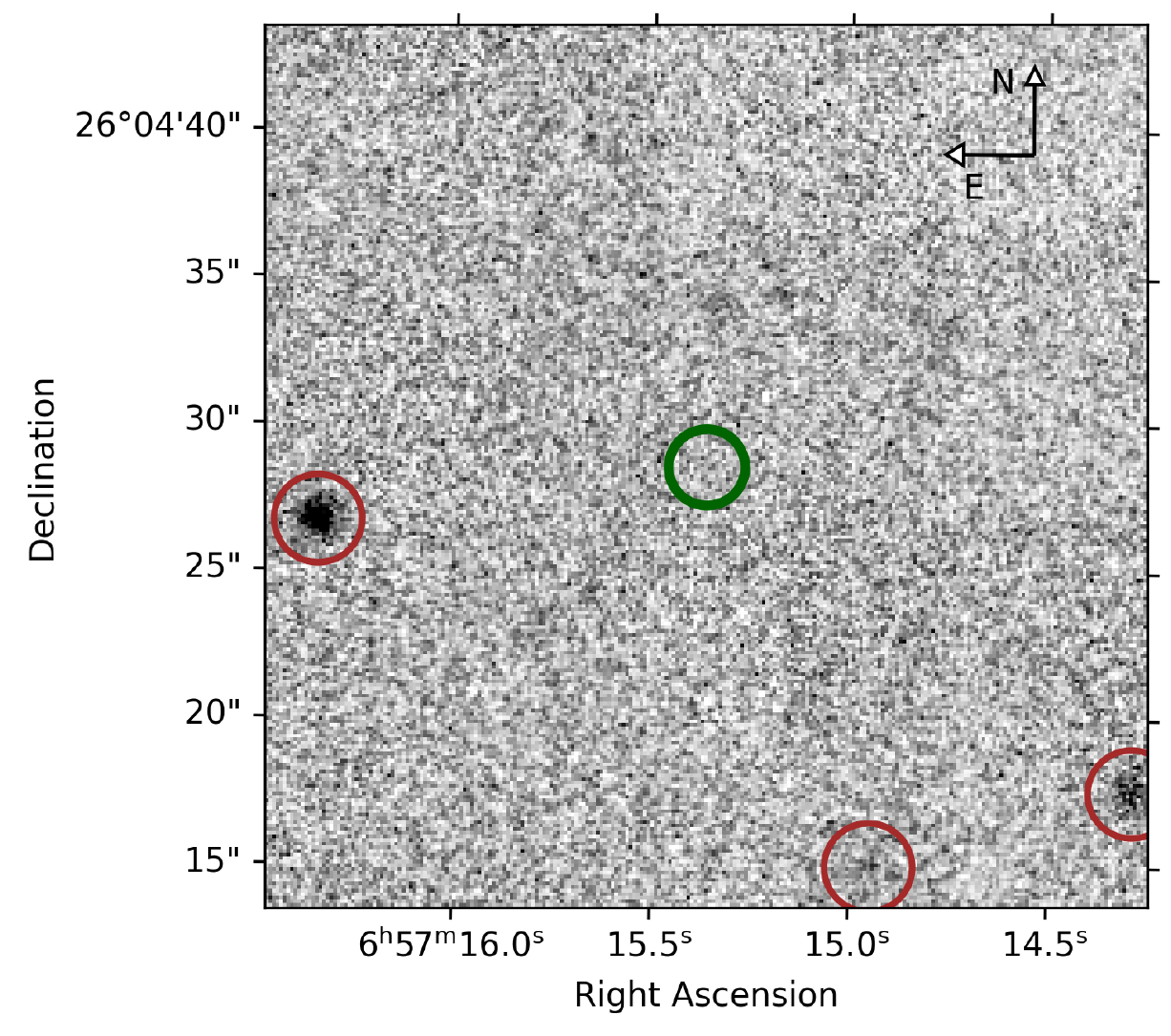}\vskip1pt
\end{center}
\end{minipage}
\begin{minipage}{0.49\textwidth}
\begin{center}
\includegraphics[width=1.01\linewidth]{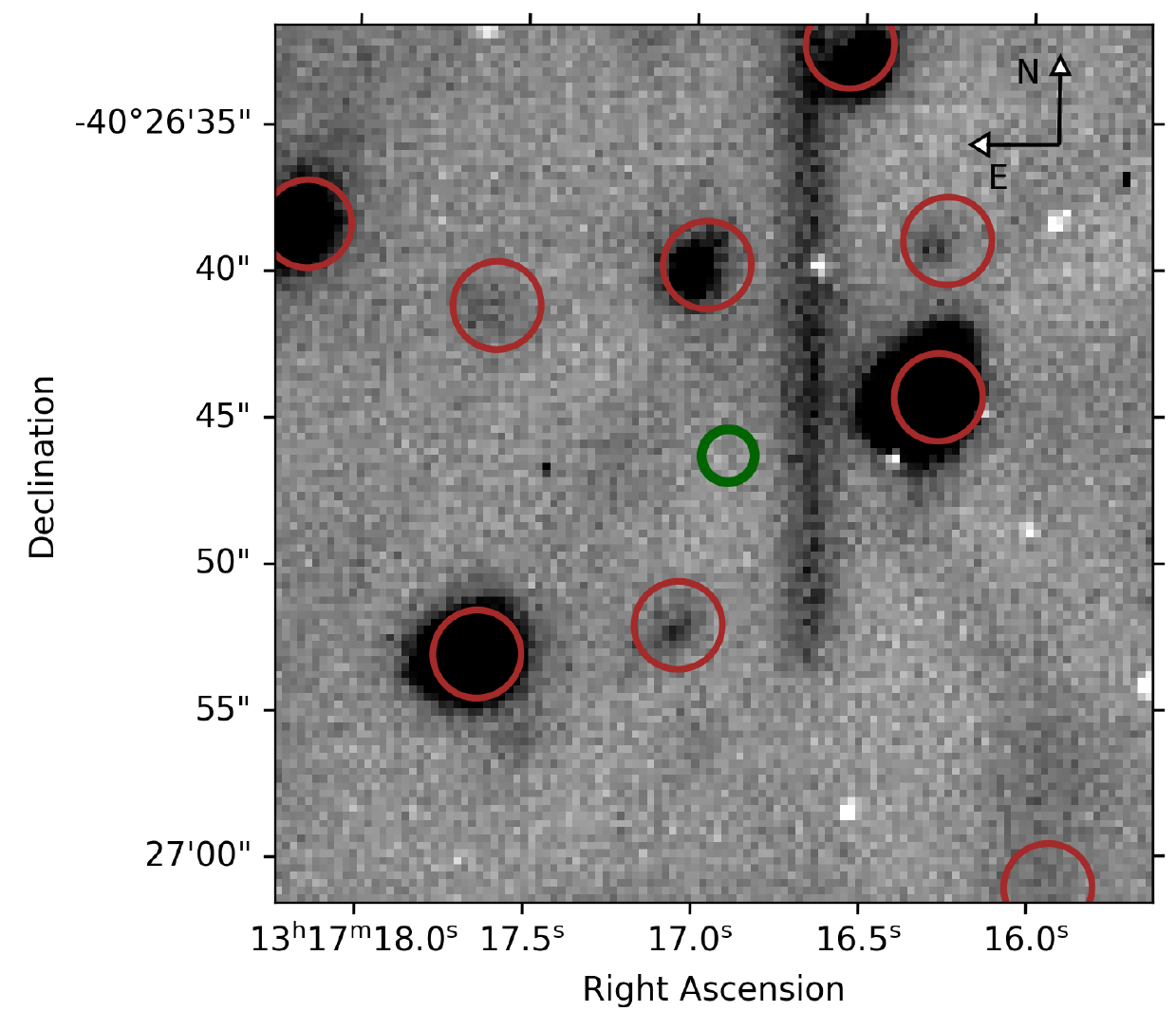}\vskip1pt
\end{center}
\end{minipage}
\caption{\lbt\ MODS $r$-band image depicting \zsfs\ (left) and \salt\ RSS image obtained for \otos\ (right). The candidate position is indicated by green circles, with the radius equivalent to the 1$\sigma$ confidence regions listed in Table~\ref{tab_erass_prop}. All nearby field sources, detected from a \texttt{SExtractor} run, are indicated with brown circles.}
\label{fig_optimgs}
\end{figure*}

\begin{figure}[t]
\begin{center}
\includegraphics[width=\linewidth]{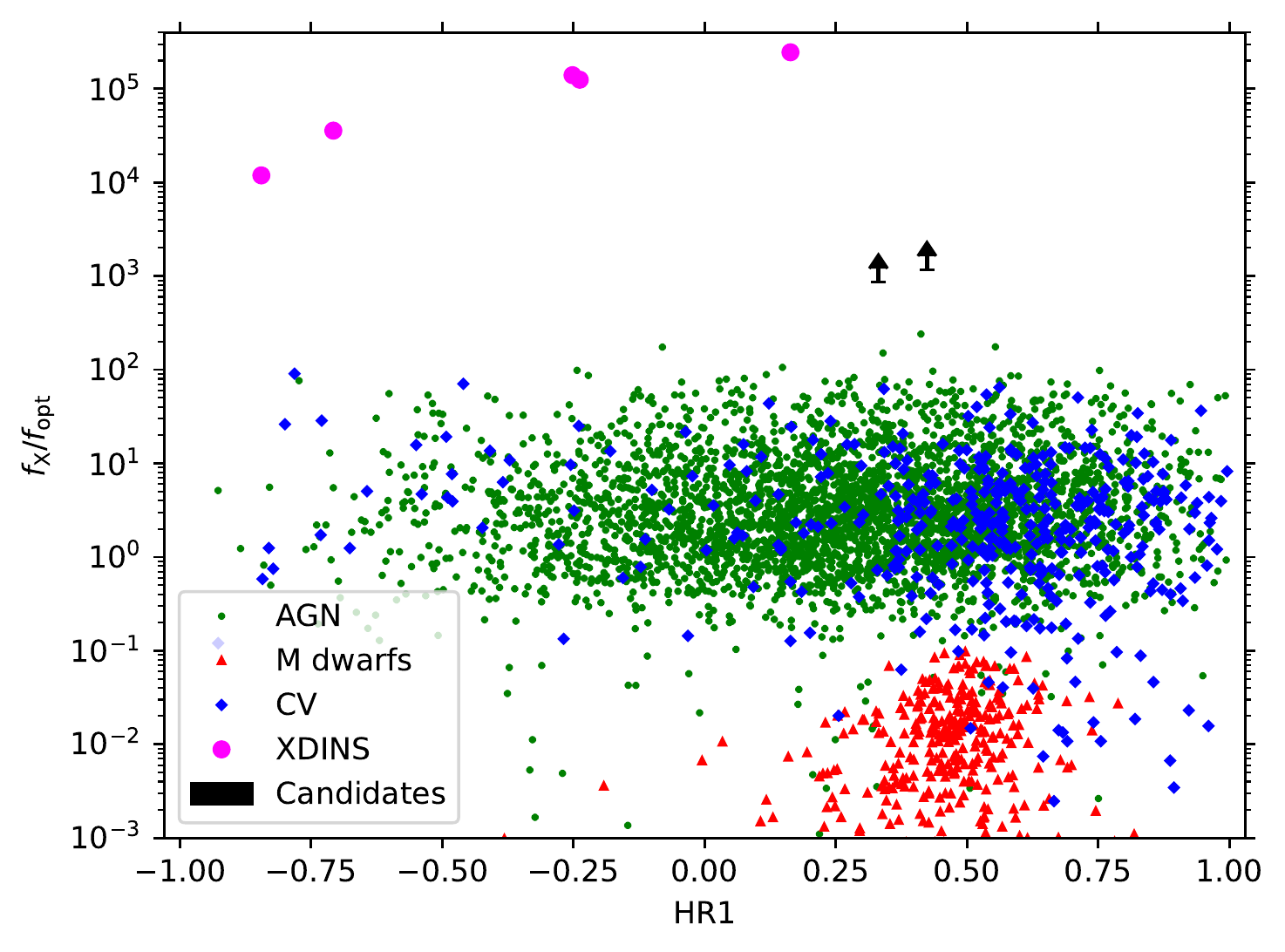}\vskip1pt
\end{center}
\caption{Candidate X-ray-to-optical flux ratio limits (black) are compared to those of the known XDINS population on the German eROSITA sky \citep[magenta, optical flux values are based on][]{2011ApJ...736..117K} and the most prevalent contaminants of our search, namely M dwarfs \citep[red, taken from ][]{2022A&A...661A..29M}, cataclysmic variables (blue, selected from the Ritter and Kolb catalogue \citep{2003A&A...404..301R} and the open CV catalogue by \citealt{2017ApJ...835...64G}), and AGN \citep[green, taken from ][]{2022A&A...661A...5L}.}
\label{fig_fxfopt}
\end{figure}

We applied the \texttt{SExtractor} software \citep{1996A&AS..117..393B}, to search for optical sources near the position of our INS candidates. The closest optical objects were detected 10$\sigma$ ($13$\arcsec\ distance, \zsfs) and 6.7$\sigma$ ($6$\arcsec\ distance, \otos) away from the respective X-ray sources' sky position. We find no viable optical counterpart in any of the INS fields (see images in Fig.~\ref{fig_optimgs}). In order to estimate the resulting X-ray-to-optical flux ratio limits, we cross calibrated the flux, measured in an optimised aperture from the \texttt{SExtractor} run, against the Pan-STARRS DR1 \citep{2016arXiv161205560C} $g$- and $r$-band (\zsfs) or the \gaia\ DR3 \citep{2022arXiv220800211G} $g$-band (\otos). The X-ray-to-optical flux ratios were computed based on the equation for the magnitude limit of a point source in an optimal Gaussian aperture (assuming a 5$\sigma$ detection limit). We determined the magnitude limit to be at magnitude 25.7 and 25.5 in the Pan-STARRS $g$- and $r$-band (\zsfs) and magnitude 25 in the \gaia\ $g$-band (\otos). The estimated photometric parameters and flux ratios are listed in Table~\ref{tab_photparam}. We compare the resulting absorbed X-ray-to-optical flux ratios to those of the most prevalent contaminants in Fig.~\ref{fig_fxfopt}. The candidates' X-ray-to-optical flux ratio values are almost a magnitude larger than those of the contaminants. Taking the interstellar absorption into account and using the Predehl \& Schmidt law \citep{1995A&A...293..889P}, as well as the reddening coefficients from \cite{1989ApJ...345..245C}, we derived unabsorbed X-ray-to-optical flux ratio limits ranging from 1000--2000.

\begin{table} [t]
\caption{Photometric parameters and magnitude limits.
\label{tab_photparam}}
\centering

\begin{tabular}{lccc}
\hline\hline
Parameter  & $g$-Band & $r$-Band & $g$-Band \\
& (\zsfs) & (\zsfs) & (\otos)\\
\hline
Magnitude Zeropoint   & 33.7 & 33.8 & 37.3\\
$\sigma_\mathrm{sky}$ & 25.1 & 35.4 & 2083\\
FWHM (px)             & 10.7 & 10.7 & 6.9\\
FWHM (\arcsec)              & 1.29 & 1.31 & 1.75\\
\hline
5$\sigma$ mag. limit  & 25.65 & 25.46 & 24.97\\
abs. $f_\mathrm{X}/f_\mathrm{opt}$  & 1023 & 859 & 1167\\
unabs. $f_\mathrm{X}/f_\mathrm{opt}$  & 1392 & 1259 & 2111\\
\hline
\end{tabular}
\end{table}


\section{Discussion and outlook\label{sec_disc}}

We report the discovery of two predominantly thermal INS candidates from eRASS. Both targets were followed up on in the optical with \lbt\ and \salt, revealing no counterparts brighter than magnitude 25. The present optical limits confidently exclude more ordinary classes of X-ray emitters than an INS \citep[][and see Fig.~\ref{fig_fxfopt}]{1999A&A...341L..51S}. The spectra of both sources are well described by simple blackbodies with $kT\sim 110$~eV, which is well in agreement for a cooling INS. The best-fit \texttt{BB} models are absorbed by column densities of $5-10 \times 10^{20}$~cm$^{-2}$, although the estimated $\nh$ exceeds the Galactic value (see Table~\ref{tab_erass_prop}) in the case of \otos. This suggests a more complex spectral shape, although we did not find a multi-component model to fit better than a single blackbody. Neutron star atmosphere models with temperatures of 30--50~eV and short distances of a few hundred parsecs do not reach the fit quality of a single \texttt{BB} component model. For both candidates, the best-fit column density is a factor of  one to two in excess of the Galactic value in the direction of the sources.

Other single-component spectral models (power law, \texttt{apec}) describe the spectra in some instances equally well. In particular, the current spectra do not allow one to discern between a strongly absorbed power law or blackbody nature. The photon indices of the best-fit power law models in Table~\ref{tab_specfit} are consistent with those of narrow-line Seyfert I galaxies observed in the \eros\ footprint \citep{2022arXiv221106184G}. However, their optical counterparts are expected to be at least 1.5 magnitudes brighter than the present optical limits we derived for our INS candidates. Likewise, a star in our Galaxy, which is a viable solution for the INS candidate \zsfs\ on the basis of its best-fit \texttt{apec} model, should be brighter than magnitude 16; this is again in stark contrast with our follow-up results. The lower temperature of the best-fit blackbody and \texttt{apec} models also disfavour a cataclysmic variable (CV) nature. While high-energy emission (above 1.5~keV) is to be expected in most CV types \citep{2017PASP..129f2001M}, we see no indication for it in the spectra of either INS candidate (see Fig.~\ref{fig_spec_plots}). Finally, for both sources we find no significant variability in either flux or spectral state over the two-year time span covered by the \eros\ data. Interestingly, the source \otos, possibly detected by \ros, may be stable over even longer timescales (30~yr). Additional observations are not available since none of the candidates have been targeted or serendipitously detected with \xmm, \cxo,\ or \swift\ in the past. We searched the radio catalogues provided in the Vizier catalogue access tool \citep{2000A&AS..143...23O}, but did not find any counterparts close to the X-ray sky positions of either candidate.

\begin{figure}[t]
\begin{center}
\includegraphics[width=\linewidth]{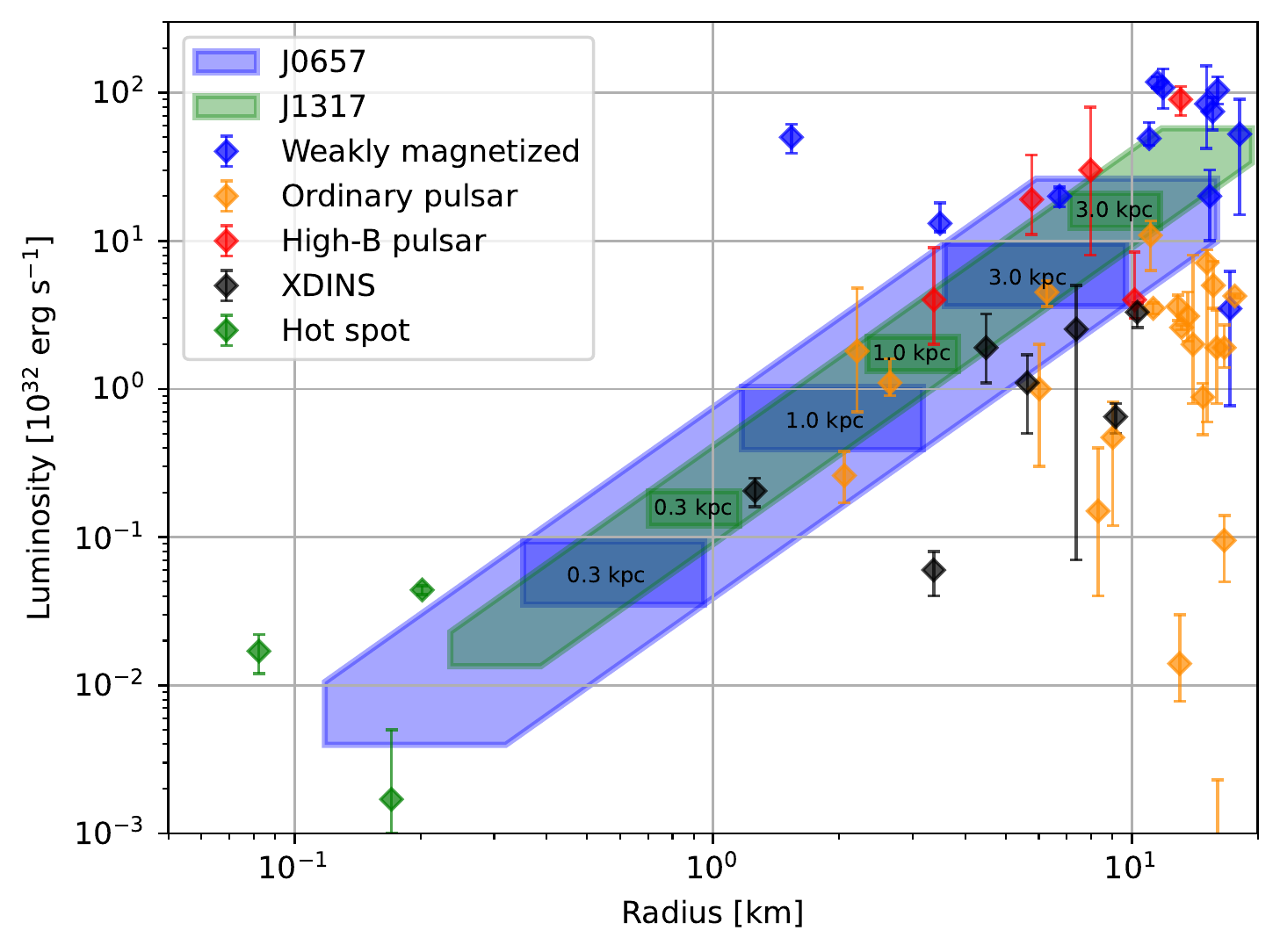}\vskip1pt
\end{center}
\caption{For \zsfs\ (blue) and \otos\ (green), we show the 1$\sigma$ confidence region distribution in luminosity and emission radius space, as it can be derived from the best-fit \texttt{BB} model depicted in Table~\ref{tab_specfit}. The rectangles indicate the 1$\sigma$ confidence region at certain distances. For the computation via the Stefan-Boltzmann law, we assumed varying test distances ranging from 0.1--5~kpc. The radius-luminosity positions of thermal emission components observed in archival INSs are marked as well. The values and object descriptions were taken from \cite{2020MNRAS.496.5052P}.}
\label{fig_lumin}
\end{figure}

To put both candidates in the context of other thermally emitting INSs, we applied the Stefan–Boltzmann law to compute luminosity estimates. Since there are no constraints on the distance, we assumed a broad range of trial distances ranging from 0.1--5~kpc. The resulting $1\sigma$ regions in the radius-luminosity diagram are marked in Fig.~\ref{fig_lumin}, along with archival cooling INS and NS hot spots collected in \cite{2020MNRAS.496.5052P}. For small distances (< 0.3~kpc), the luminosity is in agreement with a neutron star hot spot, at intermediate distances (0.3--3~kpc) the candidates could be intermediately aged rotation-powered pulsars (indicated by the ordinary pulsars in the diagram) or XDINSs, while larger distances are in agreement with a CCO-like nature or ordinary and high-B pulsars.  Based on the Galactic position and assuming the same trial distance range, the candidates are located 60--1000~pc (\zsfs) and 110--1840~pc (\otos) above the Galactic disc. Constraints on the NS age can be derived under the assumption that the candidates formed close to the Galactic plane. From the known INS population, we know that most objects possess velocities below 1000~km~s$^{-1}$ \citep{2005MNRAS.360..974H}. Based on this upper value, we computed lower age limits of 65--650~kyr (\zsfs) and 110--1100~kyr (\otos). This is in agreement with an intermediate rotation-powered pulsar or XDINS, but is at odds with a CCO nature. The reasoning is that in accordance with Fig.~\ref{fig_lumin}, weakly magnetised INSs would possess the largest distances, but this results in ages that exceed those observed for the young CCOs \citep[typical ages of a few hundred to a few thousand years, see][and references therein]{2021A&A...651A..40M}. We found magneto-thermal evolution models, taking the effects of magnetic field decay into account, as for example described in \cite{2013MNRAS.434..123V}, to give similar age estimates. We compared the luminosities to Fig.~11 in the same publication and found that the observed luminosities of $10^{31}-10^{33}$~erg~s$^{-1}$ imply ages on the 100,000 yr to a few million year timescale.

Most INSs of the classes considered above possess mixed X-ray spectra that contain both thermal and non-thermal emission components. In Sect.~\ref{sec_zsfs} and \ref{sec_otos}, we estimated the upper non-thermal to thermal flux ratio limit based on the results of a \texttt{BB+PL} fit and found those values to be at 56\% and 12\% for \zsfs\ and \otos, respectively. These ratios do not exclude the existence of a non-thermal component, since smaller ratios have also been observed for rotation-powered pulsars. For example, the rotation-powered pulsar \tmzsfs\ has a ratio of 0.3\% \citep{2005ApJ...623.1051D}, such that the apparent absence of non-thermal emission for our candidates does not allow for the possible nature  to be constrained. Additional follow-up is necessary to achieve this goal.

Both candidates conform well with an INS interpretation, but the exact nature cannot be determined on the available data alone. This may change in the near future since both candidates will be observed with \xmm\ and \nicer\ in the upcoming cycles, as part of a programme to further follow up on the most promising INS candidates from eRASS. In particular, the observations will improve the localisation and spectral parameters, as well as allow one to search for pulsations down to pulsed fractions of 5\%.

\begin{acknowledgements}
We thank the anonymous referee for thoroughly checking the manuscript and for giving suggestions and feedback that helped to improve this paper.

This work was funded by the Deutsche Forschungsgemeinschaft (DFG, German Research Foundation) – 414059771. This work was supported by the project XMM2ATHENA, which has received funding from the European Union's Horizon 2020 research and innovation programme under grant agreement n$^{\rm o}101004168$.

This work is based on data from eROSITA, the soft X-ray instrument aboard SRG, a joint Russian-German science mission supported by the Russian Space Agency (Roskosmos), in the interests of the Russian Academy of Sciences represented by its Space Research Institute (IKI), and the Deutsches Zentrum für Luft- und Raumfahrt (DLR). The SRG spacecraft was built by Lavochkin Association (NPOL) and its subcontractors, and is operated by NPOL with support from the Max Planck Institute for Extraterrestrial Physics (MPE).

The development and construction of the eROSITA X-ray instrument was led by MPE, with contributions from the Dr. Karl Remeis Observatory Bamberg \& ECAP (FAU Erlangen-Nuernberg), the University of Hamburg Observatory, the Leibniz Institute for Astrophysics Potsdam (AIP), and the Institute for Astronomy and Astrophysics of the University of Tübingen, with the support of DLR and the Max Planck Society. The Argelander Institute for Astronomy of the University of Bonn and the Ludwig Maximilians Universität Munich also participated in the science preparation for eROSITA.

The eROSITA data shown here were processed using the eSASS/NRTA software system developed by the German eROSITA consortium.

For analysing X-ray spectra, we use the analysis software BXA \citep{2014A&A...564A.125B}, which connects the nested sampling algorithm UltraNest \citep{2021JOSS....6.3001B} with the fitting environment XSPEC \citep{1996ASPC..101...17A}.

We thank S. Allanson, J. Heidt, D. Huerta, S. De Nicola, R. Saglia and D. Thompson for obtaining the LBT data. This paper uses data taken with the MODS spectrographs built with funding from NSF grant AST-9987045 and the NSF Telescope System Instrumentation Program (TSIP), with additional funds from the Ohio Board of Regents and the Ohio State University Office of Research. Some of the observations reported in this paper were obtained with the Southern African Large Telescope (SALT) under program [2021-2-LSP-001, PI: DAHB]. Polish participation in SALT is funded by grant No. MEiN nr 2021/WK/01. DAHB acknowledges research support by the National Research Foundation.

This research has made use of the VizieR catalogue access tool, CDS, Strasbourg, France (DOI : 10.26093/cds/vizier). The original description of the VizieR service was published in 2000, A\&AS 143, 23.

This work made use of Astropy:\footnote{http://www.astropy.org} a community-developed core Python package and an ecosystem of tools and resources for astronomy \citep{astropy:2013, astropy:2018, astropy:2022}. This research made use of photutils, an Astropy package for detection and photometry of astronomical sources \citep{2022zndo...6385735B}.

\end{acknowledgements}
\bibliographystyle{aa}
\bibliography{ref_j0657_j1317}

\end{document}